\documentclass[article,twocolumn,floatfix]{revtex4}

\usepackage{graphicx} 
\usepackage{subfig}
\usepackage{color}

\newcommand{\ket}[1]{\left| #1 \right\rangle}
\newcommand{\bra}[1]{\left\langle #1 \right|}

\newcommand{\be}{\begin{equation}}
\newcommand{\ee}{\end{equation}}
\newcommand{\bea}{\begin{eqnarray}}
\newcommand{\eea}{\end{eqnarray}}

\usepackage{dcolumn}
\usepackage{bm}
\usepackage{amssymb,amsmath}
\usepackage{color}
\usepackage{float}
\usepackage{tikz}
\usetikzlibrary{arrows}

\definecolor{DarkGreen}{rgb}{0,0.6,0.2}

\begin{document}
\title{Real-time emission spectrum of a hybrid atom-optomechanical cavity}
\author{Imran M. Mirza }
\affiliation{Oregon Center for Optics and Department of Physics\\University of Oregon\\
Eugene, OR 97403}
\begin{abstract}
We theoretically investigate the real-time emission spectrum of a two-level atom coupled to an optomechanical cavity (OMC). Using quantum trajectory approach we obtain the single-photon time-dependent spectrum in this hybrid system where the influence of a strong atom-cavity coupling and a strong optomechanical interaction  are studied. We find a dressed state picture can explain the spectra by predicting the exact peak locations as well as the relative peak heights. In our analysis we also include the effect of mechanical losses (under weak mechanical damping limit) and single-photon loss through spontaneous emission from the two-level emitter. 
\end{abstract}
\maketitle
\section{Introduction}
 Hybrid atom-optomechanical quantum systems have shown a significant amount of research activity in last few years \cite{schliesser2011hybrid, rogers2014hybrid, wallquist2010single, restrepo2014single}. A mixture of two well developed fields of cavity quantum electrodynamics (CQED)\cite{haroche1989cavity, kavokin2007microcavities} and quantum optomechanics \cite{kippenberg2008cavity, aspelmeyer2014cavity}, these hybrid systems are anticipated to be an integral part of future quantum technologies where a wealth of phenomena can arise due to atom-light and light-mechanics interactions at the quantum level. Recently, it has been theoretically proposed and experimentally observed that these hybrid systems in which a mechanical oscillator interacts simultaneously with a radiation field and with a matter-like system (a single or a multi-atom system e.g. a Bose-Einstein condensate), can be used to implement many interesting effects of pure quantum nature. Some current and compelling examples include: Optomechanical cooling by $\wedge$-configured atoms \cite{bariani2014hybrid}, entanglement generation and detection in atom-optomechnical systems \cite{de2011entanglement}, quantum state control of Rydberg atoms using optomechanics \cite{bariani2014single} and generation of macro-level quantum superposition of nano-particles using spin-mechanical interactions in NV diamond centers inside nano-diamonds \cite{yin2013large}.\\
A relevant and important question in the context of such hybrid systems is what type of spectrum is emitted by such structures and what can be inferred from the spectrum about different features of atom-cavity and cavity-mechanics coherent interactions? In this paper we answer this question by theoretically exploring the single-photon time-dependent spectrum emitted by an atom-OMC system in the strong atom-cavity and strong optomechnaical interaction regimes (hereafter called the strong-strong coupling case). The experimental realization of a strong coupling at the single-photon level in a cavity QED setup has already been demonstrated  in past \cite{kimble1998strong}, while achieving single-photon strong optomechanical coupling in an OMC is anticipated to be realized in near future (as reflected by a substantial progress on the theoretical as well as experimental side in the single-photon optomechnaics \cite{brooks2012non, akram2010single, nunnenkamp2011single, liao2012spectrum}). \\
Recently we have investigated the time-dependent spectrum of a single-photon initially trapped in an empty OMC cavity \cite{mirza2014single}. Here we extend that study to a more interesting situation when the actual single-photon source (an excited two-level emitter) is coupled to the OMC. This problem (from the perspective  of standard CQED setups) is more appealing as the observation of vacuum Rabi splitting \cite{hennessy2007quantum} in the single-photon CQED is a key indicator of achieving strong coupling between a two-level atom and an optical cavity \cite{cui2006emission}. In present study we explore the modifications strong optomechanical interaction brings into the Rabi-splitted spectrum and more importantly we would like to answer that how these modifications emerge in the spectrum as a function of time? Here it is worthwhile to mention that there are some studies conducted in recent past, in which a laser or single photon driven hybrid atom-OMC system's spectrum is calculated in the stationary limit (time-independent version) \cite{breyer2012light,jia2013single, jacobs2012probe}. But here we focus on the time-dependent/real-time version of the spectrum, which to our knowledge has not be reported yet in the present context.\\
We find that even working in the strong atom-cavity and strong optomechanical coupling regimes (which are two necessary conditions to observe vacuum Rabi splitting in the CQED setups and multiple side-band peaks in the context of cavity optomechanics, respectively), it takes finite time ($t=20\omega^{-1}_{M}$) to observe the fully resolved spectrum in an atom-OMC system. Moreover, spectrum evolves from a broad Lorentzian to a frequency doublet and finally additional side-band appears in the spectrum. Our analysis gives quantitative values of time when these spectral changes can be observed. The time order in which spectrum grows and manifests these features indicate different physical process responsible for the production of multiple resonances in the spectrum.\\
The paper is organized as follows: In Sec.~II we present our theoretical model describing the hybrid atom-optomechanical cavity setup. Next in Sec.~III we discuss our results in detail, focusing on the single-photon spectrum emitted in real time (time-dependent spectrum) when there is a strong coupling between atom and the cavity mode and also between single photon and movable cavity mirror in the OMC. A dressed state picture is employed to understand the spectrum by exactly locating the spectrum resonances and relative peak heights. Furthermore the effect of phonon losses (in the waek mechanical damping limit) and the events of atomic spontaneous emission are also examined. Finally in Sec.~IV we summarize the conclusions of this work. \\

\section{Theoretical description}
\subsection{Model and Hamiltonian}
\begin{figure*}[t]
\hspace{-20mm}\includegraphics[width=4.4in]{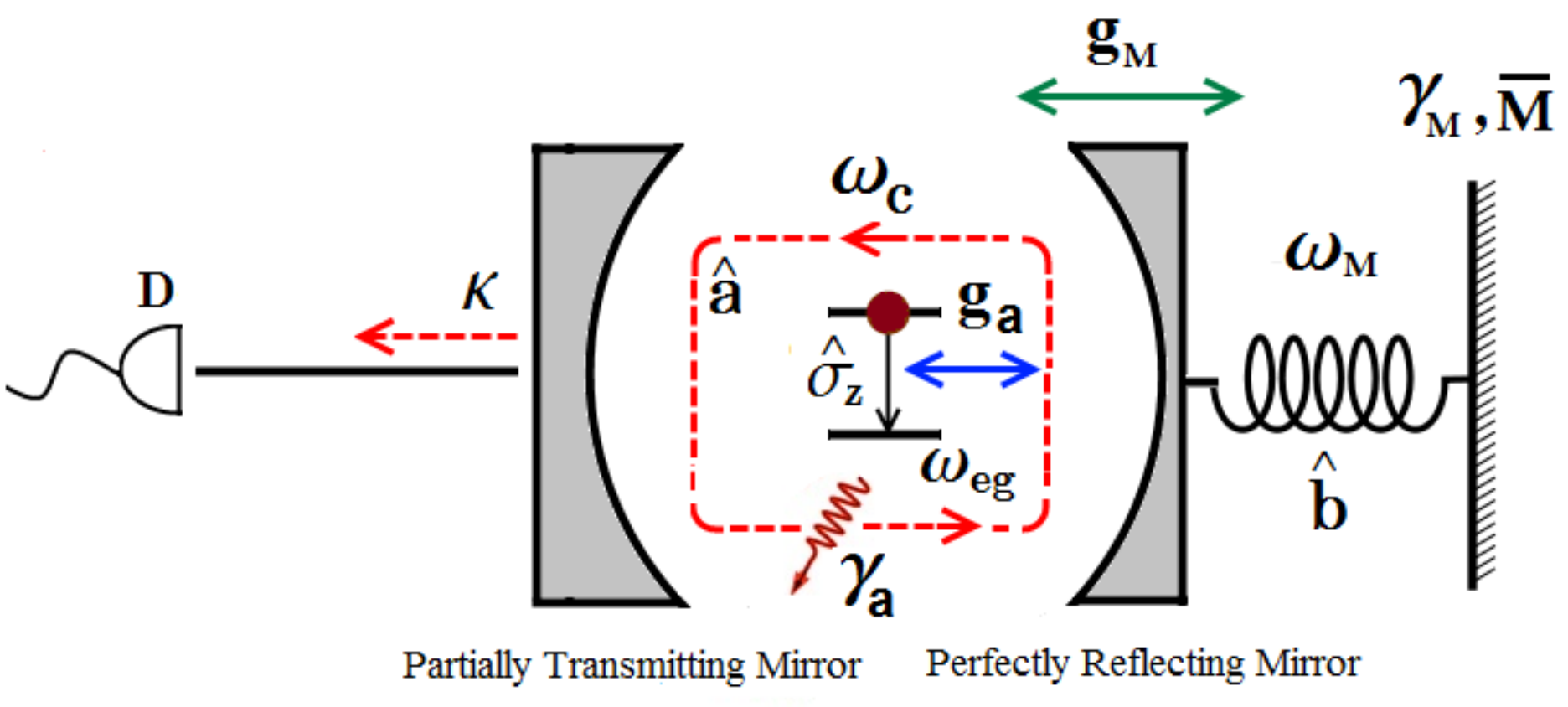}
\captionsetup{
  format=plain,
  margin=1em,
  justification=raggedright,
  singlelinecheck=false
}
 \caption{(Color online) An initially excited two-level atom in an OMC. Initially the OMC is assumed to have zero photons and phonons. The optical field emitted by OMC is guided through an optical fiber to the detector $D$ and its time-dependent spectrum is used to probe the signatures of atom-cavity and optomechanical interaction dynamics. For further deatils about the model of the system see Sec.~II(A).}\label{Fig1}
\end{figure*}
As shown in Fig.~1, system under consideration consists of an initially excited two level atom embedded in an optomechnical cavity. This two level atom has a ground state ($\ket{g}$) and an excited state ($\ket{e}$) with transition frequency and spontaneous emission rate given by $\omega_{eg}$ and $\gamma_{a}$ respectively. Experimentally such a two level system can be realized using a solid state quantum dot \cite{hennessy2007quantum} or an NV center in diamond \cite{barclay2009coherent} or an artificial atom in the context of circuit quantum electrodynamics (CirQED) \cite{you2011atomic}. For the sake of simplicity atom's external degrees of freedom are neglected and optomechnaical cavity is taken to be a usual Fabry-Per\'ot cavity with a fixed but partially transmitting (movable but perfectly reflecting) mirror on left (right) side. Optical cavity is assumed to have a single isolated resonant mode given by frequency $\omega_{c}$ while the annihilation of photons in the cavity is described by the action of operator $\hat{a}$. The atom-field interaction strength is given by the rate $g_{a}$.\\
We have considered the movable mirror to be a quantum harmonic oscillator with equilibrium frequency $\omega_{M}$ and the destruction of quanta of vibration (phonons) is represented in term of an operator $\hat{b}$. The parameter $g_{M}$ is used to indicate the optomechanical interaction strength. The mechanical oscillator is coupled with a heat bath as well which causes phonons to decay at a rate $\gamma_{M}$ and the bath temperature is described in terms of average phonon number $\overline{M}$. The OMC can also leak out photons  at a rate $\kappa$ into a dispersion-less optical fiber which is assumed to have a continuum of modes. Optical fiber guides the single-photon wavepacket such that finally it can be detected at the output detector $D$. Neglecting the optical and mechanical baths, the system Hamiltonian under rotating wave approximation and in a frame rotating with frequency $\omega_{c}$ can be expressed as:
\begin{equation}
\begin{split}
&\hat{H}_{sys}= -\hbar\Delta_{a}\hat{\sigma}_{-}\hat{\sigma}_{+}+\hbar(g_{a}\hat{a}^{\dagger}\hat{\sigma}_{-}+g^{\ast}_{a}\hat{a}\hat{\sigma}_{+})+\hbar\omega_{M}\hat{b}^{\dagger}\hat{b}\\
&\hspace{10mm}-\hbar g_{M}\hat{a}^{\dagger}\hat{a}(\hat{b}^{\dagger}+\hat{b})
\end{split}
\end{equation}
Here $\Delta_{a}=\omega_{eg}-\omega_{c}$ and we have neglected the zero point energies of the optical and mechanical oscillators. Non-vanishing commutation relations are given by: $[\hat{\sigma}_{+},\hat{\sigma}_{-}]=\hat{\sigma}_{z}$, $[\hat{a},\hat{a}^{\dagger}]=1$ and $[\hat{b},\hat{b}^{\dagger}]=1$. Here $\hat{\sigma}_{z}\equiv \ket{e}\bra{e}-\ket{g}\bra{g}$ while $\hat{\sigma}_{+}/\hat{\sigma}_{-}$ are the atomic raising/lowering operators and $\hat{\sigma}_{z}$ is the usual Pauli operator. On the right side of Eq.~(1), the first term describes the atomic Hamiltonian, second term denote the atom-cavity coupling, third term the Hamiltonian of a free quantum mirror and finally the fifth term represents the non-linear optomechanical coupling.\\
\subsection{Quantum trajectory method}
To study the dynamics of such a hybrid atom-optomechnaical system when it is coupled to fiber's mode continuum, we employ the quantum trajectory/jump approach (QJA) \cite{dalibard1992wave, carmichael1999statistical, van2000quantum}. This open quantum system method is known to be an appropriate technique as it incorporates both the presence of photo-detectors in the setup as well as different sources of photon and phonons losses. In QJA the registration of an excitation (either a photon or phonon) is described by a jump operator. In present situation we have three types of jumps possible: optical jumps (${\rm \hat{J}^{(O)}_{out}}$) which describe a situation in which a photon leaks out from the cavity into the fiber and detected at detector ${\rm D}$, atomic jumps (${\rm \hat{J}^{(A)}_{out}}$) through spontaneous emission mechanism and mechanical jumps (${\rm \hat{J}^{(M)}_{out}}$) representing phononic-decays. Theses stochastically occurring jump events are described by the following set of operators:
\begin{subequations}
\begin{eqnarray}
\hat{J}^{(O)}=\sqrt{\kappa}\hat{a},\\
\hat{J}^{(A)}=\sqrt{\gamma_{a}}\hat{\sigma}_{-},\\
\hat{J}^{(M)}=\sqrt{\gamma_{M}}\hat{b},
\end{eqnarray}
\end{subequations}
According to QJA, in the absence of quantum jumps, system evolves according to a non-Unitary Schr\"odinger equation which is given by:
\begin{equation}
i\hbar\frac{d}{dt}\ket{\tilde{\psi}(t)}=\hat{H}_{NH}\ket{\tilde{\psi}(t)}
\end{equation}
where $\hat{H}_{NH}$ is a non-Hermitian Hamiltonian constructed from the system Hamiltonian $\hat{H}_{sys}$ (which is Hermitian) and an anti-Hermitian Hamiltonian constructed from the jump operators as follows:
\begin{equation}
\hat{H}_{NH}=\hat{H}_{sys}-\frac{i\hbar}{2}\hat{J}^{\dagger(O)}\hat{J}^{(O)}-\frac{i\hbar}{2}\hat{J}^{\dagger(A)}\hat{J}^{(A)}-\frac{i\hbar}{2}\hat{J}^{\dagger(M)}\hat{J}^{(M)}
\end{equation} 
Note that in writing above Hamiltonian we have assumed the mechanical heat bath to be at zero temperature  (i.e. $\overline{M}=0$). The ket $\ket{\tilde{\psi}(t)}$ is the so called no-jump state which is constructed from all different possibilities of finding the excitations in the system before being leaked out. In present case no-jump state takes the following form:
\begin{equation}
\ket{\tilde{\psi}(t)}=\sum_{m=0}^{\infty}a_{m}(t)\ket{e,0,m}+\sum_{m=0}^{\infty}b_{m}(t)\ket{g,1,m}
\end{equation}
 on the right hand side of above equation we are following the notational convention that the first slot in the ket labels atomic state, second slot indicates the number of photons in the optical cavity and third slot represents the number of phonons in the mechanical oscillator. In present paper we'll focus on a situation when there are neither photons nor phonons present in the system initially i.e. initial no-jump state is $\ket{e,0,0}$. Now by applying the machinery of QJA, in the next section, we'll calculate the time-dependent spectrum.
\subsection{Time-dependent spectrum}
To understand the growth of spectrum as a function of time, in this section we calculate the single-photon time-dependent spectrum produced by an atom-OMC system. For this calculation we utilize a setup in which the single photon wavepacket emitted by the atom-OMC systems passes through a filter cavity which has a fixed bandwidth $\Gamma$ but a variable frequency $\omega$ and then after leaving this filter cavity the photon is registered by detector D. The time-dependent spectrum is then interpreted as the filtered counting rate after many repetitions of the same experiment and it can be expressed by using the definition first introduced by Eberly and Wodkiewicz (E\&W) \cite{eberly1977time} as given by:
\begin{equation}
\begin{split}
&N(t;\Delta,\Gamma)=\Gamma^{2}\int_{0} ^t\int_{0} ^t e^{-(\Gamma-i\Delta)(t-t^{'})}e^{-(\Gamma+i\Delta)(t-t^{''})}\times\\
&\hspace{20mm}\langle\hat{a}^{\dagger}(t^{'})\hat{a}(t^{''})\rangle dt^{'}dt^{''}.
\end{split}
\end{equation}
where $\Delta=\omega-\omega_{c}$ and in the actual E\&W spectrum definition we have replaced the classical field amplitudes by the corresponding optical field annihilation and creation operators. Sometime to make comparison with the already reported stationary spectra, (as give by the Wiener Khinchin theorem \cite{cui2006emission}) above spectrum is also integrated over the entire time range. We'll call this integrated spectrum as the synthesized spectrum $N_{s}(t;\Delta,\Gamma)$ and in present article we'll always report this synthesized version of the spectrum.
\section{Results and discussion}
There are many parameters involved in the dynamics of the system under study, we here study the effect of varying three parameters on the single-photon time-dependent spectrum. These parameters represent different loss mechanisms and are: cavity decay rate $\kappa$, mechanical damping rate $\gamma_{M}$ and spontaneous emission rate $\gamma_{a}$. We'll start with the simplest situation when the atomic spontaneous emission and mechanical damping are completely neglected (i.e. we are setting $\gamma_{M}=0$, $\overline{M}=0$ and $\gamma_{a}=0$ for Sec.~III(A)). 
\subsection{Absence of mechanical and spontaneous emission losses}
\begin{figure*}[t]
\centering
  \begin{tabular}{@{}cccc@{}}
    \includegraphics[width=2.25in, height=1.7in]{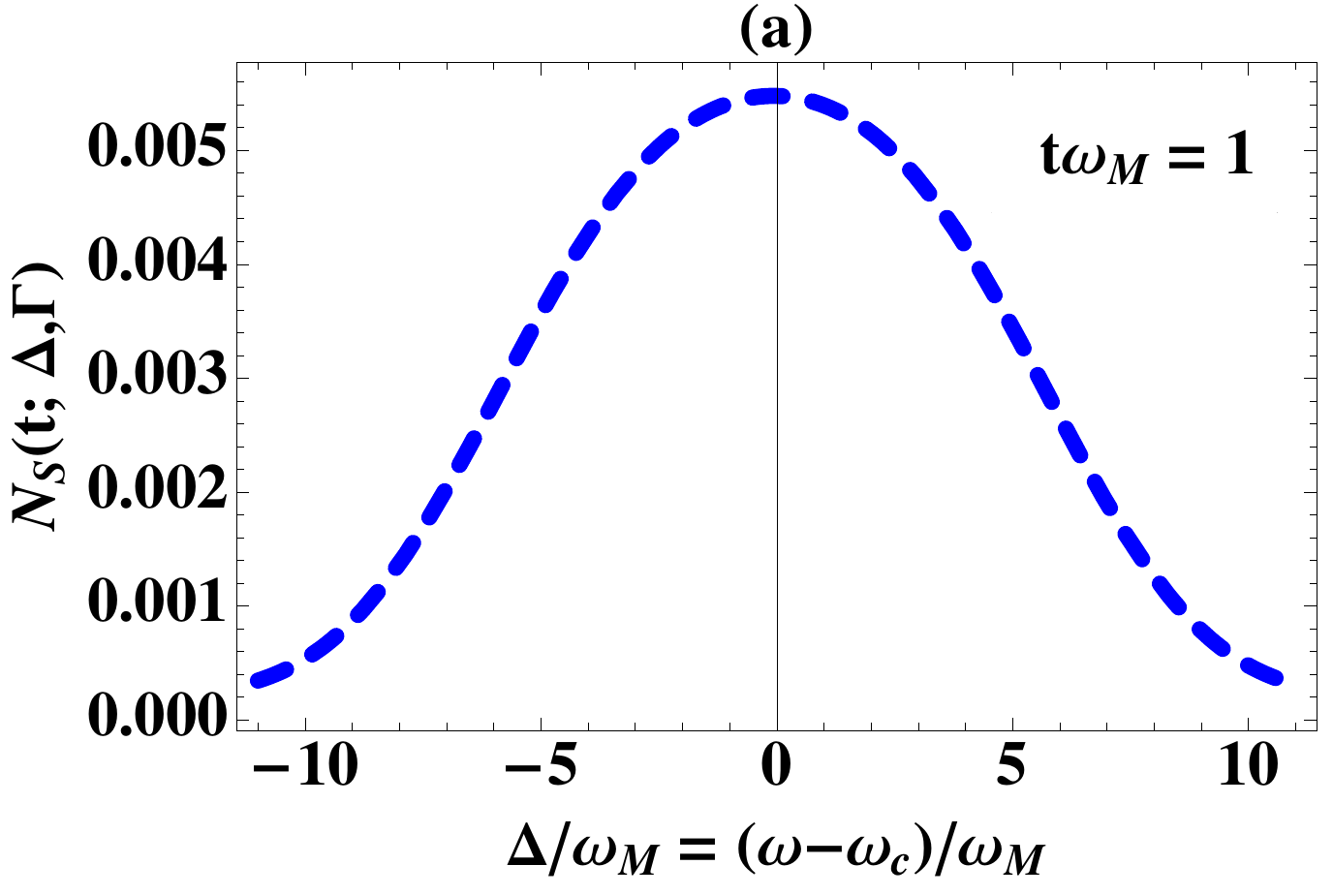} &
    \includegraphics[width=2.25in, height=1.77in]{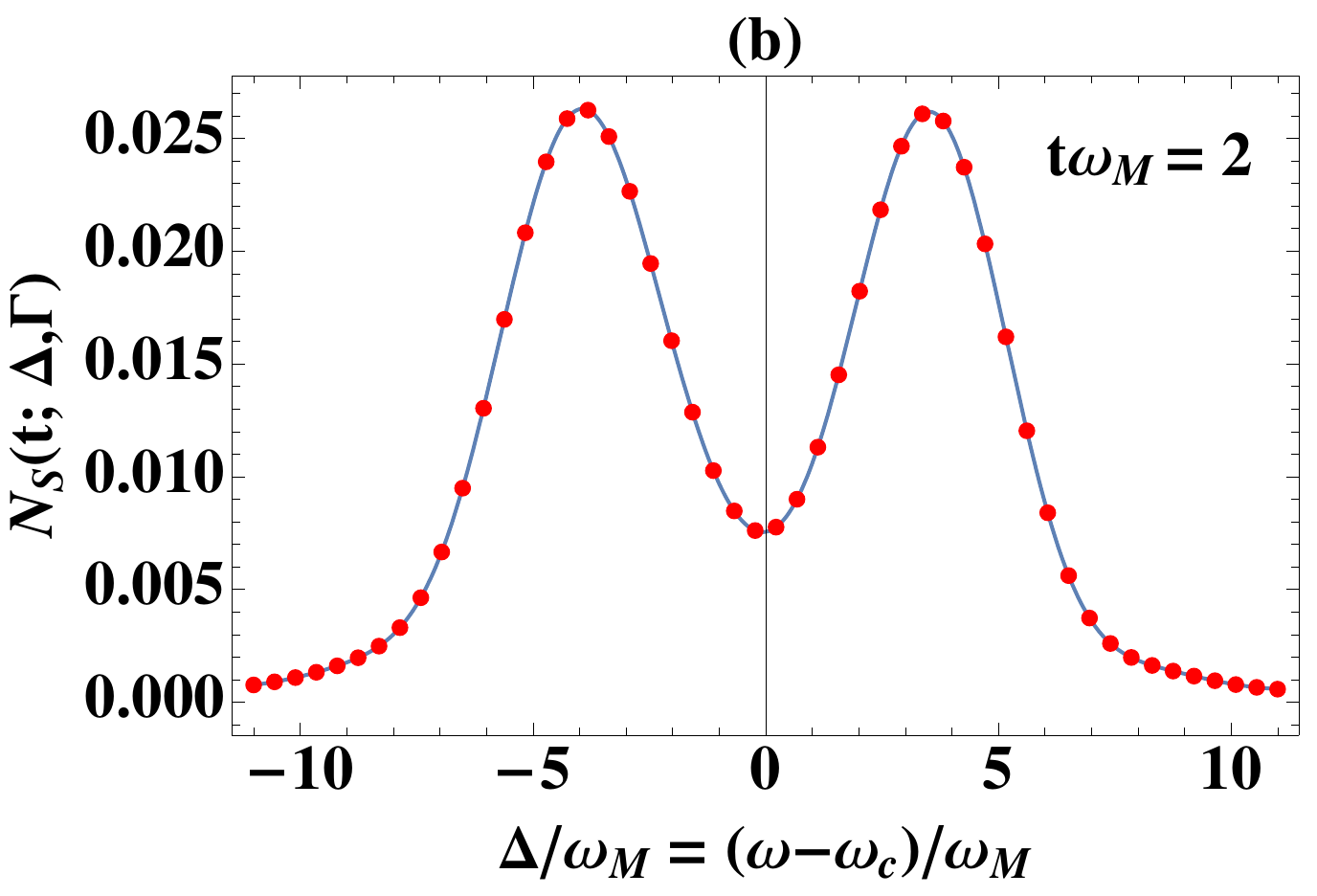} &
    \includegraphics[width=2.25in, height=1.7in]{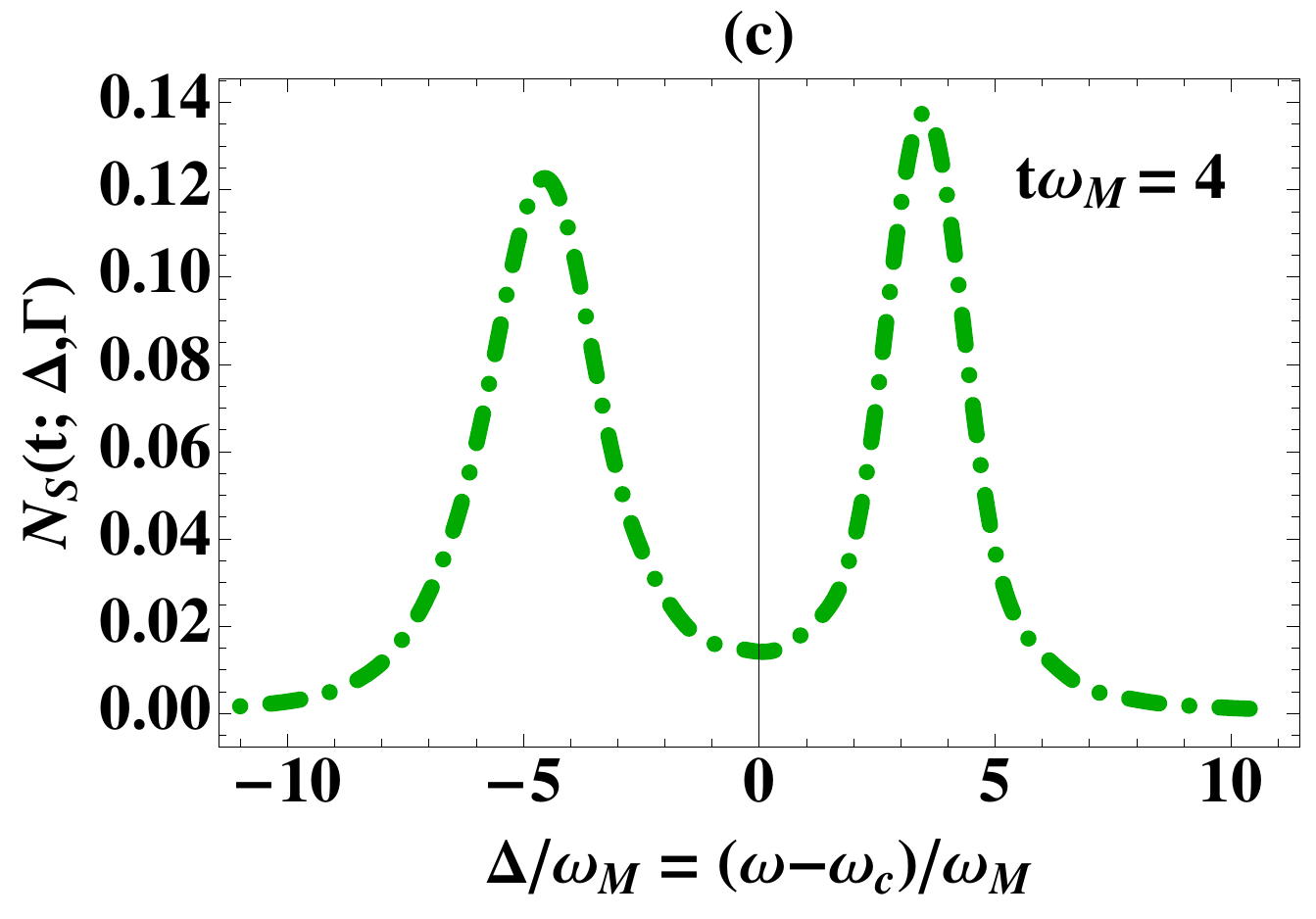} \\
    \includegraphics[width=2.1in, height=1.72in]{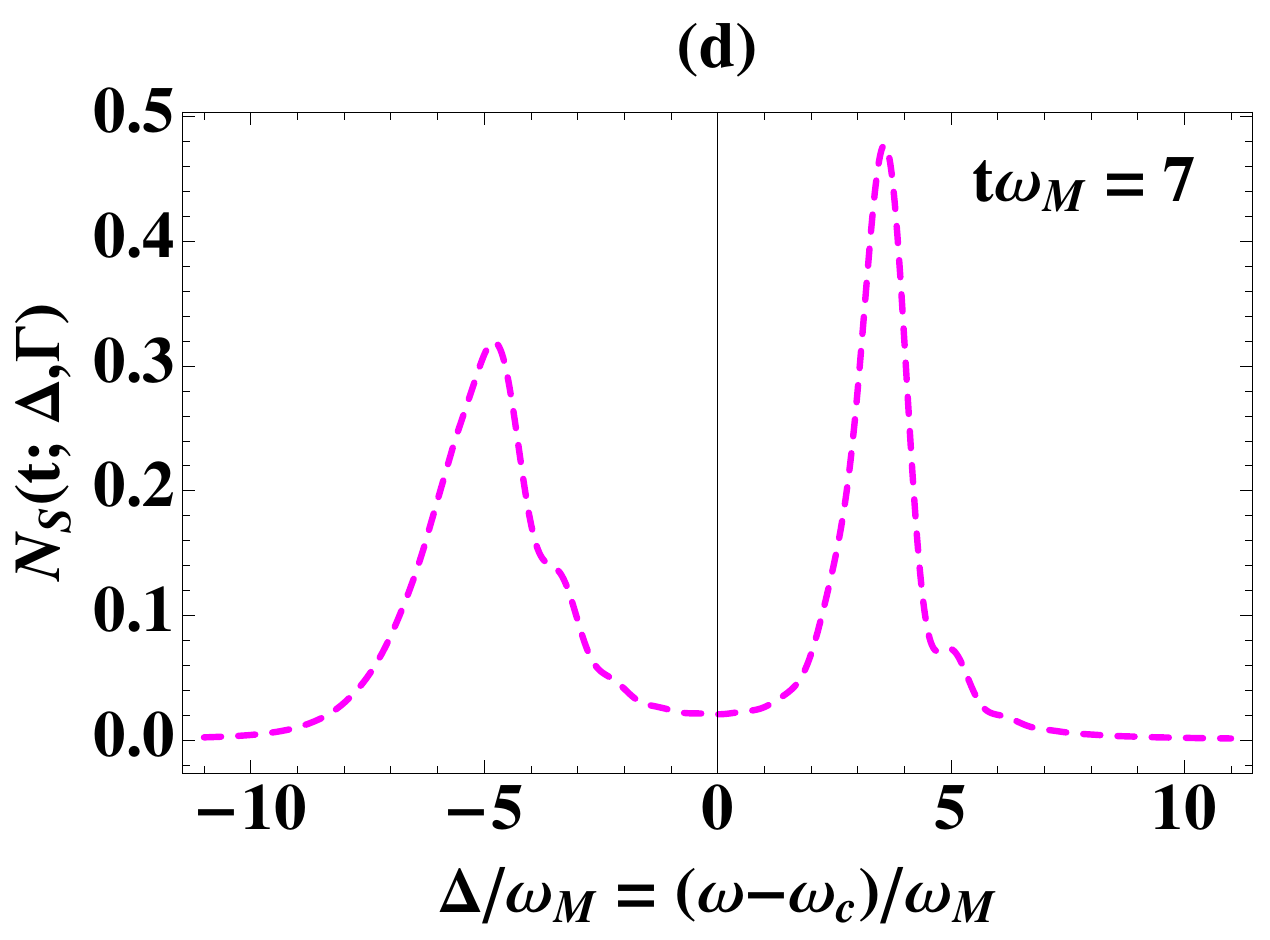} &
    \hspace{2mm}\includegraphics[width=2.1in, height=1.7in]{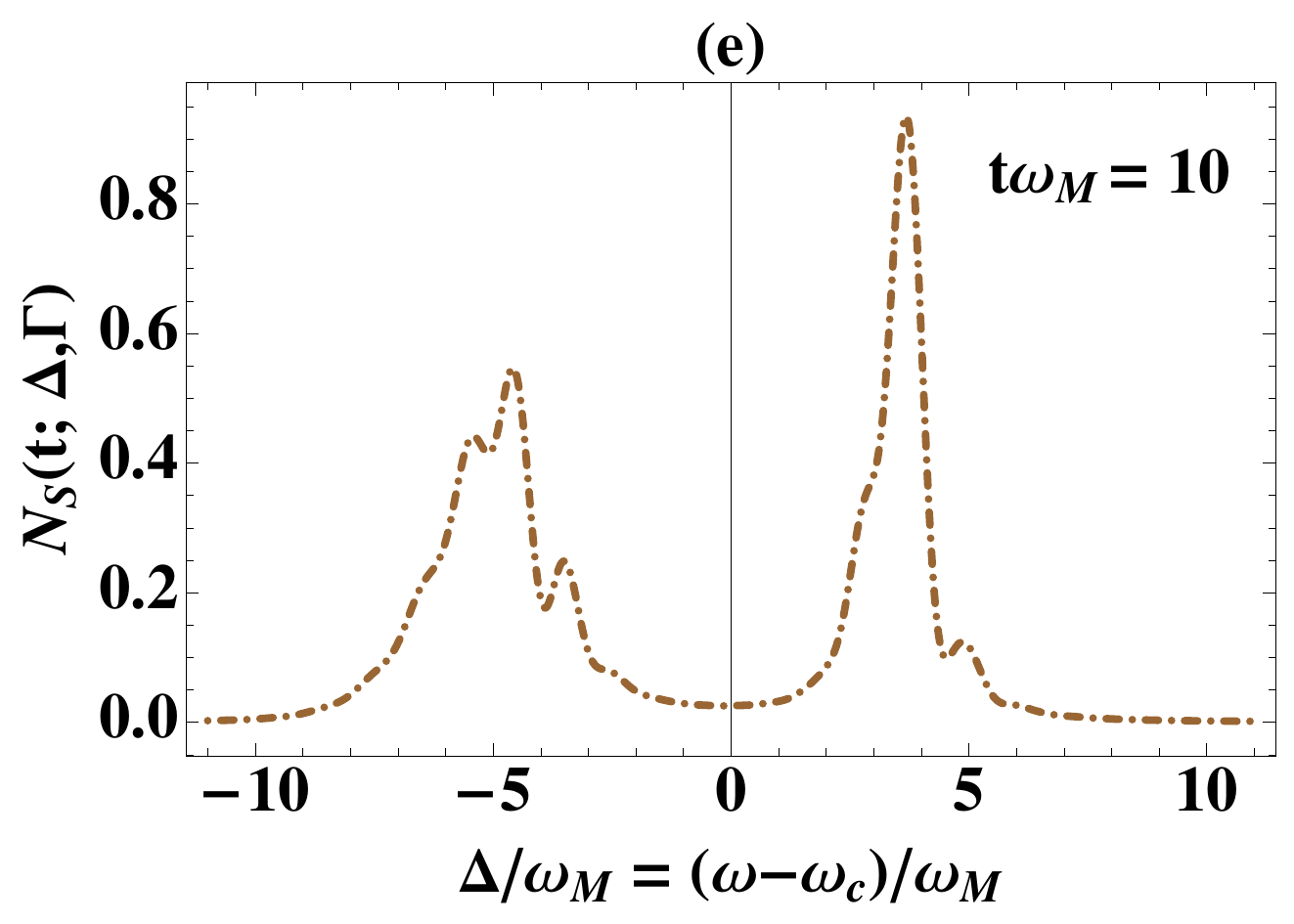} &
    \hspace{2mm}\includegraphics[width=2.1in, height=1.7in]{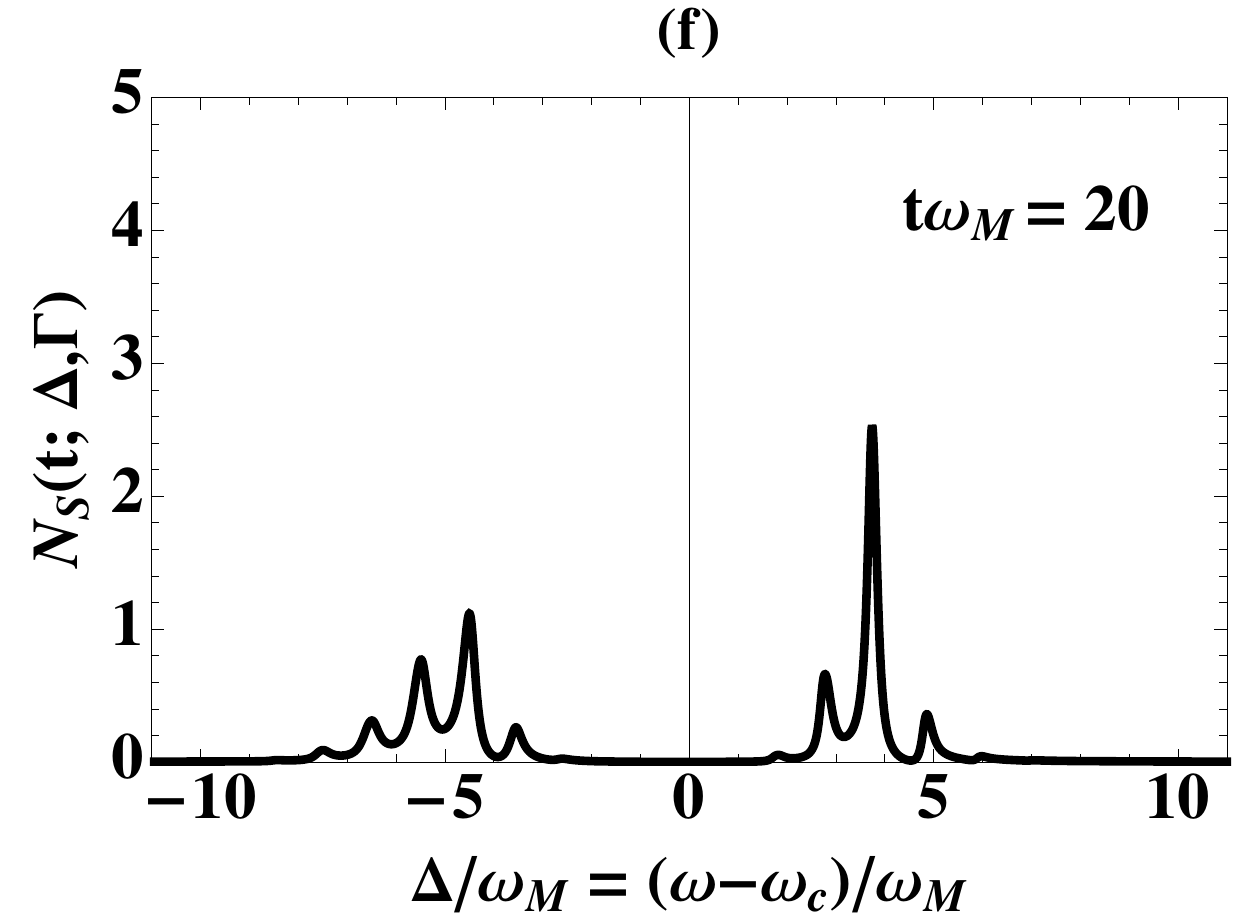} 
  \end{tabular}
  \captionsetup{
  format=plain,
  margin=1em,
  justification=raggedright,
  singlelinecheck=false
}
  \caption{(Color online) Single-photon time-dependent spectrum in a hybrid atom-OMC system, calculated at six different times, in the strong-strong coupling regime and within good cavity limit. $t = 1\omega^{-1}_{M}, 2\omega^{-1}_{M}, 4\omega^{-1}_{M}$, $7\omega^{-1}_{M}, 10\omega^{-1}_{M}$ and $20\omega^{-1}_{M}$ correspond to thick blue dashed, red dotted, green dotted dashed, thin magenta dashed, brown thin dotted dashed and black solid curves respectively. We have considered an on-resonance case i.e. $\Delta_{a}=0$ while other parameters used are: $g_{a}/\omega_{M} = 4$, $g_{M}/\omega_{M} = 1.2$, $\kappa/\omega_{M} = 0.5$,  $\Gamma/ \omega_{M} =0.1 $. Notice that spectrum starts as a broad Lorentzian and then vacuum Rabi-spitting appears in the spectrum, finally blue and red side-bands emerge almost at the same time, which later on become more visible in the spectrum. In above plots, we have considered the possibility of having upto ten phonons in the mechanical motion. This choice is made noticing that with taking higher number of phonons ($m > 10$) the overall structure of the spectrum doesn't change rather only more tiny side-bands appear at higher $\pm\Delta/ \omega_{M}$ values which lie outside the range of the $\Delta/\omega_{M}$-axis shown.}
\end{figure*}
In this subsection we'll assume that the cavity leakage rate is the main and dominant source of photon decay which can be realized experimentally by assuming that $\kappa\gg \gamma_{a}$ and phonons decay is completely ignored (i.e. $\gamma_{M} = 0$). Under these conditions time-dependent spectrum takes the following form:
\begin{equation}
\begin{split}
&N(t;\Delta,\Gamma)=\kappa\Gamma^{2}\int_{0} ^t e^{-\Gamma t^{'}}\Bigg|\sum_{m=0}^{\infty}\int_{0}^{t^{'}}e^{(i\Delta+\Gamma/2+im\omega_{M})t^{''}}\\
&\times b_{m}(t^{''})dt^{''}\Bigg|^{2}dt^{'}.
\end{split}
\end{equation}
 In Fig.~2 we have plotted the single-photon time-dependent spectrum in the strong-strong coupling case ($g_{a}/\kappa=8$, $g_{M}/\kappa=2.4$). We have chosen this regime because it is well known that under this choice of parameters an atom-cavity system and an empty OMC system, both separately show a fully resolved spectra \cite{liao2012spectrum, nunnenkamp2011single}. Note that in the case of OMC spectrum one has to respect the additional good cavity limit ($\omega_{M}>\kappa $) as well. \\
Staring from Fig.~2(a), we notice that in the beginning ($t\omega_{M}=1$) spectrum appears as a broad Lorentzian curve and it takes at least $t=2\omega^{-1}_{M}$ time for the spectrum to split into two peaks. This is an indication that the spectrum has started to exhibit vacuum Rabi splitting. But it takes further $ 2\omega^{-1}_{M}$ time for the spectrum to show full Rabi splitting i.e. both peaks in the strong atom-coupling regime get separated by a frequency equal to $2g_{a}$ (as shown in Fig. 2(c) and as one expects from the standard Rabi CQED model). This specifies a physical process in which an initially excited atom emits a  photon into the cavity mode which is exchanged between the atom and cavity few times before leaking out from the cavity (see Fig.~2(c)). Note that we have taken $g_{a}>g_{M}$ thus even if the single photon undergoes Rabi oscillations before being leaked out from the cavity and produces full Rabi-splitting yet the optomechanical interaction doesn't leave its strong signatures on the spectrum.\\
Around $t=7\omega^{-1}_{M}$, we begin to notice optomechanical interaction imprints on the spectrum in the form of additional side-bands (tiny shoulders). These side-bands are showing that now the single-photon has interacted for long enough time with the movable mirror such that it can now produce observable effects in the spectrum. Note also that both red and blue side-bands are visible on the negative and positive $\Delta/\omega_{M}$-axis respectively. Physically red (blue) side-bands are produced when single photon interacts with the movable mirror and loses (gains) some energy from the mechanical oscillator, thus coming out from the cavity with lower (higher) frequency as discussed in detail in Ref.~\cite{ren2013single,liao2012spectrum}. \\
In the case of single-photon time-dependent spectrum emitted form an empty OMC (as reported in our recent work \cite{mirza2014single}) the red side-bands should appear in the time-dependent spectrum before the emergence of blue side-bands. But here we notice that both side-bands appear simultaneously. This shows that in the process of generating fully splitted vacuum Rabi spectrum photon has interacted with the mechanical mirror but due to small optomechanical interaction compared to $g_{a}$ and due to broadening of vacuum Rabi spectrum peaks, red side-bands are not resolved before the blue ones. At later times (as shown in Fig.~2(e) at $t=10\omega^{-1}_{M}$), due to optomechnaical and atom-cavity interactions occurring for longer times, we find both of the Rabi peaks become sharper as well as both blue and red side-bands start to show up simultaneously at integer multiple of $\omega_{M}$ apart from the main Rabi peaks. Finally at $t=20\omega^{-1}_{M}$ we obtain the stationary limit of the spectrum as also reported in \cite{jia2013single}. It is worthwhile to note that in Ref. \cite{jia2013single} this stationary spectrum is calculated based on a real-space quantification approach as first introduced by Shen et.al  in 2009 \cite{shen2009theory}. Here, along with reporting the time-dependent version of the spectrum, we have also provided an alternate method of calculating the time-independent spectrum using the quantum trajectory method, which gives the exact same results as one can find by applying other methods for example the method of Shen et.al.\\

In addition to the conclusions made in references \cite{jia2013single} (to which our stationary spectrum results are in agreement), we conclude from the time-dependent spectrum that even if we are working in a strong-strong coupling regime still it takes $\sim t = 20\omega^{-1}_{M}$ time to observe fully resolved spectrum. Moreover, the time-dependent spectrum also reveals the order in which different resonances appear in the spectrum describing the dynamic role played by different physical mechanisms which are responsible for the production of these spectral peaks.
Finally at the end of this section, we would like to mention that by taking $g_{a}=0$ and $g_{M}=0$ in the stationary spectrum one can recover the usual single-photon spectra obtained in the context of CQED \cite{cui2006emission} and an empty OMC \cite{liao2012spectrum} respectively.

\subsubsection{Dressed State Analysis and Asymmetry in peak heights}
\begin{figure*}[t]
\includegraphics[width=6in, height=3in]{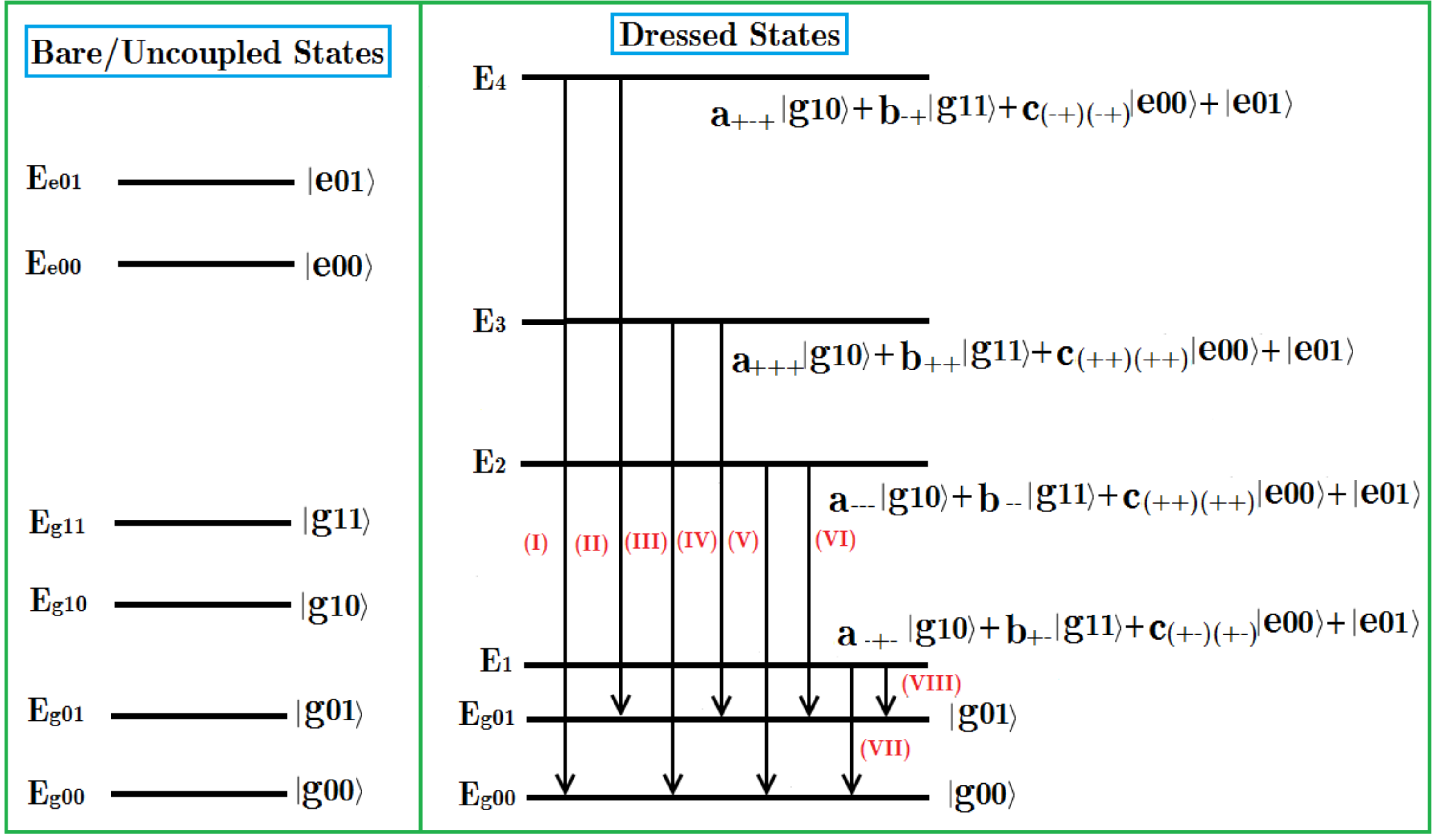}
\captionsetup{
  format=plain,
  margin=1em,
  justification=raggedright,
  singlelinecheck=false
}
\captionsetup{
format=plain,
margin=1em,
justification=raggedright,
singlelinecheck=false
}
\caption{(Color online) Dressed state picture of a hybrid atom-OMC system. Only single photon and single phonon restricted energy level diagram is shown. We notice that due to atom-cavity coupling and optomechanical interaction upper four states in the bare state picture are mixed up to form four dressed states. Transition from these excited states to two ground states are responsible for the production of eight resonances as shown in Fig.~4. Moreover this dressed state picture can also be used to explain the peak asymmetry present in Fig.~4 (see text for details).}
\end{figure*}
In order to understand the stationary/time-independent spectrum we now present a dressed state analysis of the problem. For that we'll consider a simple case when the atom-OMC starts off with atom in the excited state and movable mirror in its quantum ground state while later on there can be at most a single phonon present in the system. Note that this single phonon restriction is basically an artificial situation because the condition $\frac{g_{M}}{\kappa}<< 1$ should be satisfied for the single-phonon approximation to be valid in an actual experiment. But the imposition of the condition $\frac{g_{M}}{\kappa}< 1$ is equivalent to working in the weak optomechanical coupling limit in which a fully resolved spectrum can't be observed. Yet for simplicity (and to present analytic results) we'll report this analysis here which can easily be extended (numerically) to as many as ten phonons case while in that situation the assumption of strong optomechnaical coupling is going to be a valid.\\

To perform a dressed state analysis we neglect all source of decoherence and concentrate on the system Hamiltonian (as given in Eq.~1). With a single photon and a single phonon restriction one can define the basis: $\lbrace \ket{g,0,0},\ket{g,0,1}, \ket{g,1,0},\ket{g,1,1},\ket{e,0,0},\ket{e,0,1} \rbrace$ needed for the diagonalization of the system Hamiltonian. As a result we obtain the following set of eigenvalues  (in unrotated frame) with different coefficients appearing in the un-normalized eigenvectors shown in Fig.~3 given by:
\begin{widetext}
\begin{subequations}
\begin{eqnarray}
\Bigg\lbrace  E_{g00},E_{g01},E_{\pm\pm} \Bigg\rbrace = \Bigg\lbrace \hbar\omega_{g},\hbar\omega_{g}+\hbar\omega_{M},\frac{\hbar}{2}(\omega_{M}\pm\sqrt{4g^{2}_{a}+2g^{2}_{M}+\omega^{2}_{M}\pm 2\sqrt{4g^{2}_{a}g^{2}_{M}+g^{4}_{M}+4g^{2}_{a}\omega^{2}_{M}}  })  \Bigg\rbrace\\
a_{\pm\pm\pm}=\frac{-g^{2}_{M}+\omega^{2}_{M}\pm\sqrt{4g^{2}_{a}g^{2}_{M}+g^{4}_{M}+4g^{2}_{a}\omega^{2}_{M}}  \pm\omega_{M}\sqrt{4g^{2}_{a}+2g^{2}_{M}+\omega^{2}_{M}\pm 2\sqrt{4g^{2}_{a}g^{2}_{M}+g^{4}_{M}+4g^{2}_{a}\omega^{2}_{M}}  }}{2g_{a}g_{M}}\\
b_{\pm\pm}=\frac{-\omega_{M}\pm\sqrt{4g^{2}_{a}+2g^{2}_{M}+\omega^{2}_{M}\pm 2\sqrt{4g^{2}_{a}g^{2}_{M}+g^{4}_{M}+4g^{2}_{a}\omega^{2}_{M}}  }}{2g_{a}}\\
c_{(\pm\pm)(\pm\pm)}=\Bigg( \frac{-2(g^{3}_{a}-(\frac{g_{a}\omega_{M}}{2}\pm\frac{g_{a}}{2}\sqrt{4g^{2}_{a}+2g^{2}_{M}+\omega^{2}_{M}\pm 2\sqrt{4g^{2}_{a}g^{2}_{M}+g^{4}_{M}+4g^{2}_{a}\omega^{2}_{M}}  }    ))}{\frac{-g_{a}g_{M}\omega_{M}}{2}\pm g_{a}g_{M}\sqrt{4g^{2}_{a}+2g^{2}_{M}+\omega^{2}_{M}\pm 2\sqrt{4g^{2}_{a}g^{2}_{M}+g^{4}_{M}+4g^{2}_{a}\omega^{2}_{M}}  }}     \Bigg)\times\\ \nonumber
 \hspace{20mm}\Bigg(\omega_{M}\pm\frac{1}{2}(-\omega_{M}\pm\sqrt{4g^{2}_{a}+2g^{2}_{M}+\omega^{2}_{M}\pm 2\sqrt{4g^{2}_{a}g^{2}_{M}+g^{4}_{M}+4g^{2}_{a}\omega^{2}_{M}}  })\Bigg)
\end{eqnarray}
\end{subequations}
\end{widetext}
while in Fig.~3  we have $E_{-+}=E_{1},E_{--}=E_{2},E_{+-}=E_{3},E_{++}=E_{4}$. We note that due to atom-cavity and optomechanical interaction four bare states get mixed to form four dresses states. From the upper (excited) four dressed states system can make a photonic transition to two ground states. We note that this process gives rise to a total of eight transitions and these transition frequencies exactly matches the peak locations obtained in the spectrum as shown in Fig.~4. Here we would like to indicate that there exists an alternate method of calculating peak locations as well. This method is performed by setting the real part of the poles equal to zero in the final expression of the time-independent spectrum \cite{liao2012spectrum}. We confirmed that the peak locations obtained from both methods (dressed state analysis and the pole-method) gives us identical results. \\
It is worthwhile to mention here that the eigenvalues obtained upto single phonon are not in agreement with the result obtained in the reference \citep{jia2013single} (energy eigenvalues vary as $\sim \hbar\omega^{-1}_{M}\sqrt{g^{4}_{M}+4 g^{2}_{a}\omega^{2}_{M}}$). In this reference (and some other references \cite{nunnenkamp2011single, liao2012spectrum} related to the subject of single photon optomechanics) a polaron transformation is applied to transform the non-linear optomechanical cavity into a frame where the Hamiltonian can easily be digonalized.  The reason for this disagreement again lies on our artificial single phonon restriction. We checked (numerically) that when more and more phonons are added in our model both results (one which is obtained by following our quantum trajectory method and the other which performs a polaron transformation) start to match. Thus this issue can easily be resolved by extending the present single phonon restricted analysis to a multiple phonon situation. \\
\begin{figure}
\includegraphics[width=2.8in, height=2.1in]{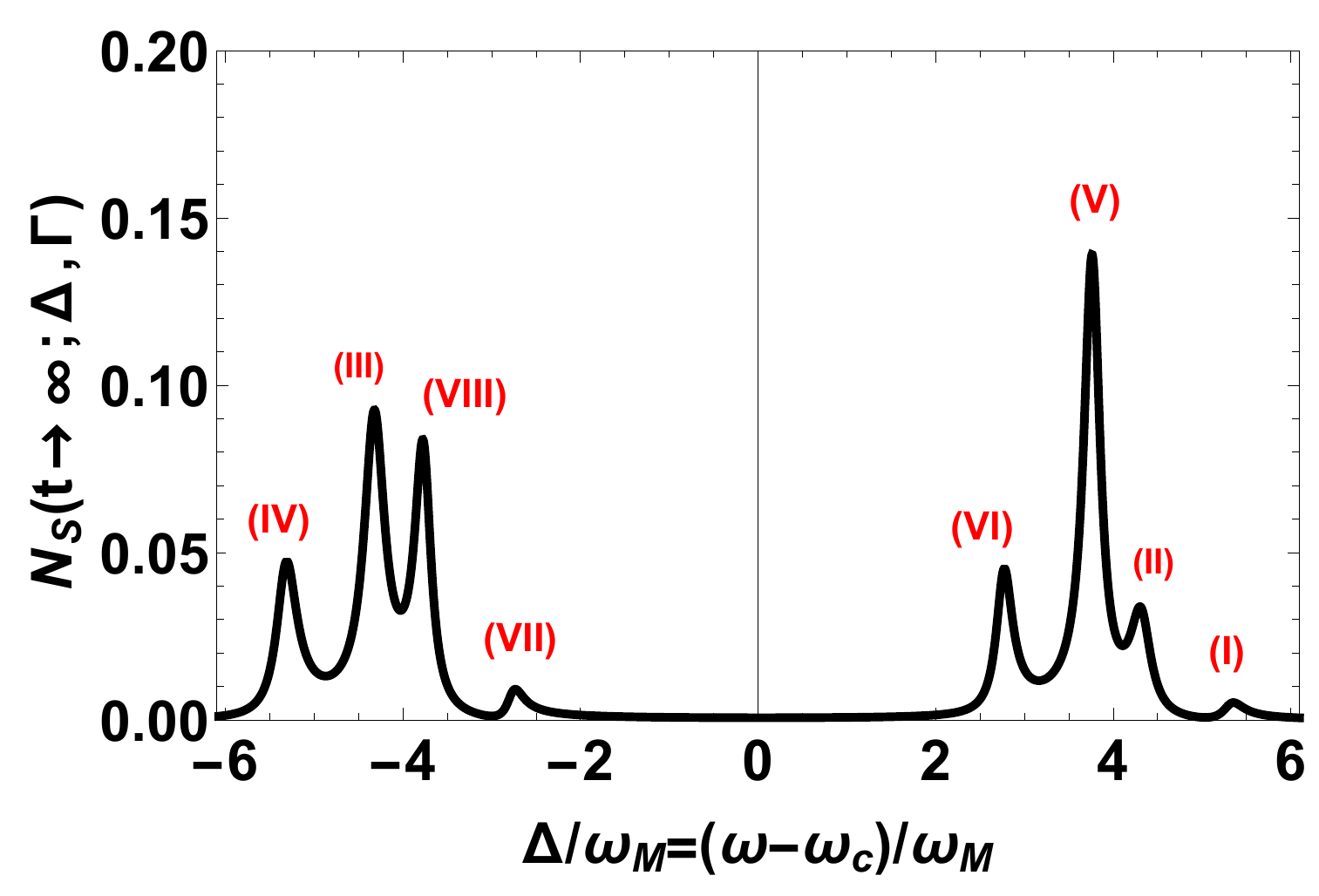}
\captionsetup{
  format=plain,
  margin=1em,
  justification=raggedright,
  singlelinecheck=false
}
\captionsetup{
format=plain,
margin=1em,
justification=raggedright,
singlelinecheck=false
}
\caption{(Color online) Single-photon single-phonon stationary spectrum emitted by an atom-OMC system. This spectrum is plotted to confirm that the dressed state analysis shown in Fig.~3 provides exact peak locations. Different peak labels corresponds to different transitions among dressed states as marked in Fig.~3. Parameters are the same as used in Fig.~2.}
\end{figure}
Another advantage of performing a dressed state analysis is that it also explains the asymmetry in the peak heights which is not possible to explain otherwise (for example by by setting the real part of pole equal to zero). In the dressed state picture we find that with the presence of single photon in the atom-OMC, system can be in any one of the four upper/excited states. In general, the state of the system at a later time ${\rm t}$ will be in a superposition of all four excited states and since the form of superposition can be imbalanced hence the transition from any one of these excited states to bottom ground states ($\ket{g00},\ket{g01}$) will happen with unequal probability thus causing an asymmetry in the peak heights. Finally we would like to remark that we numerically extended this dressed state analysis to the case of ten phonons case (corresponding spectrum shown in Fig.~2(f)) and we found that dressed state method gives exact peak locations as well as describe the unequal peak heights for that case too (the analysis not shown here).

\section{Inclusion of Mechanical Damping and Spontaneous Emission}
\begin{figure*}[t]
\centering
  \begin{tabular}{@{}cccc@{}}
    \includegraphics[width=2.25in, height=1.7in]{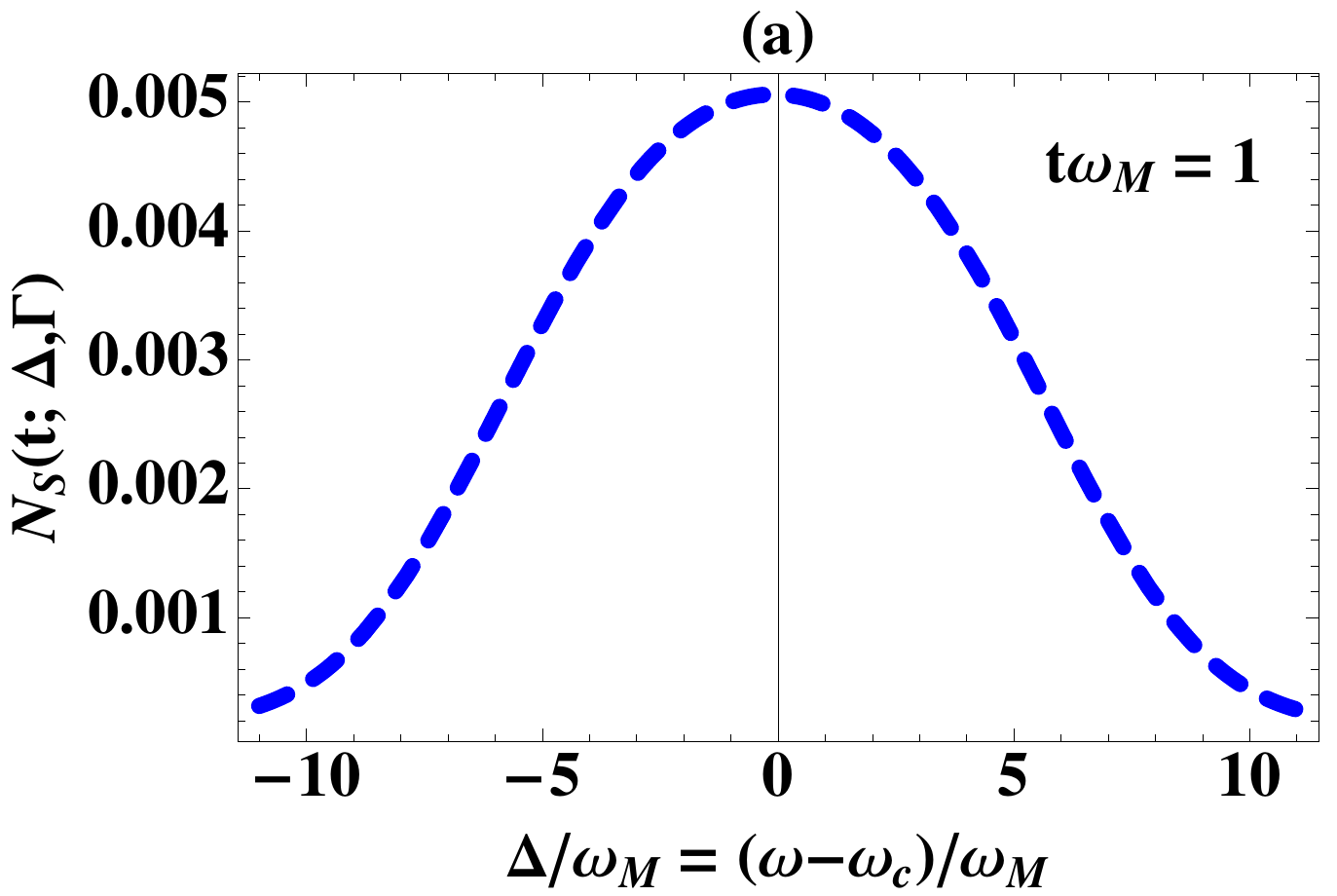} &
    \includegraphics[width=2.25in, height=1.7in]{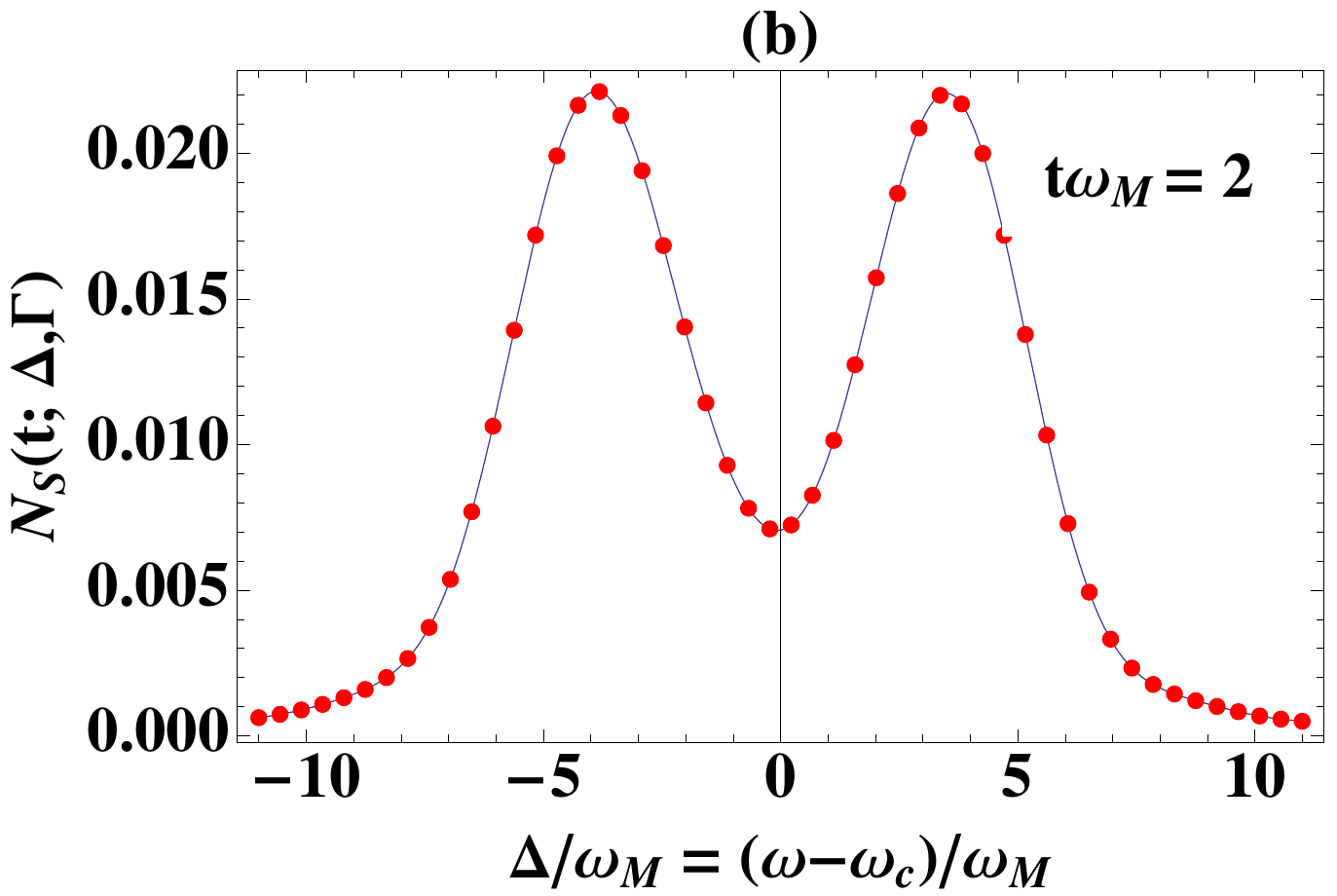} &
    \includegraphics[width=2.25in, height=1.7in]{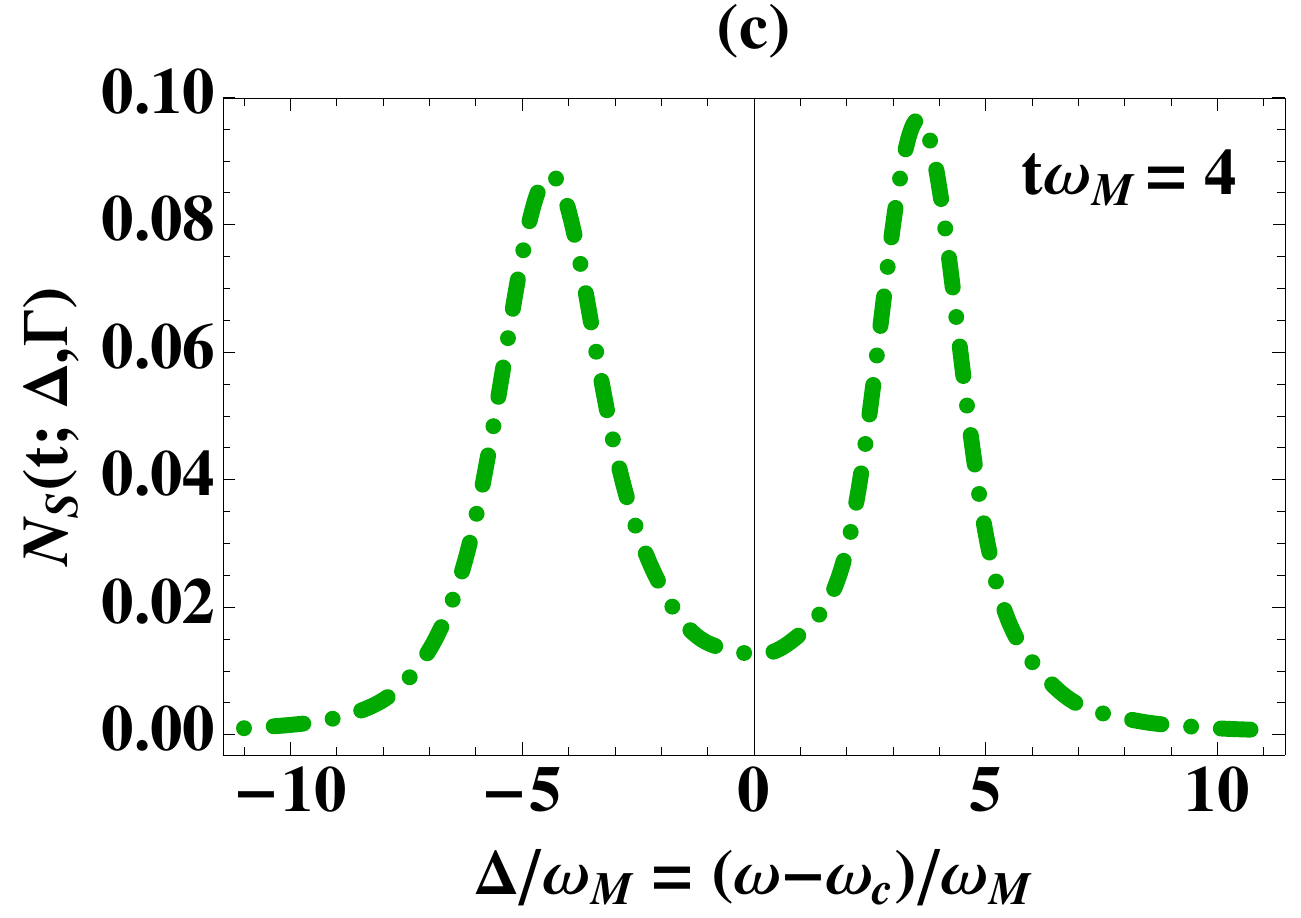} \\
    \includegraphics[width=2.25in, height=1.7in]{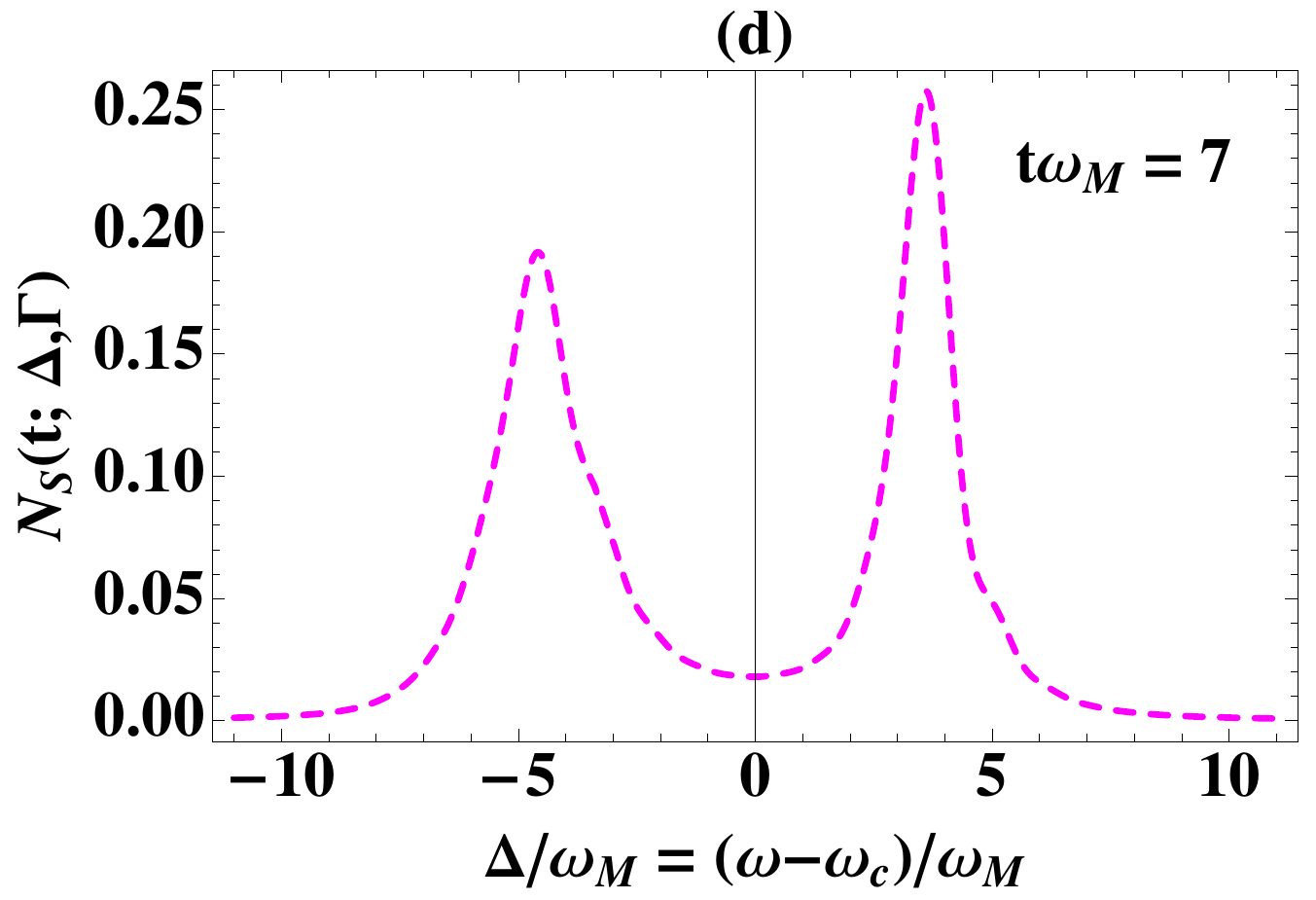} &
    \includegraphics[width=2.25in, height=1.7in]{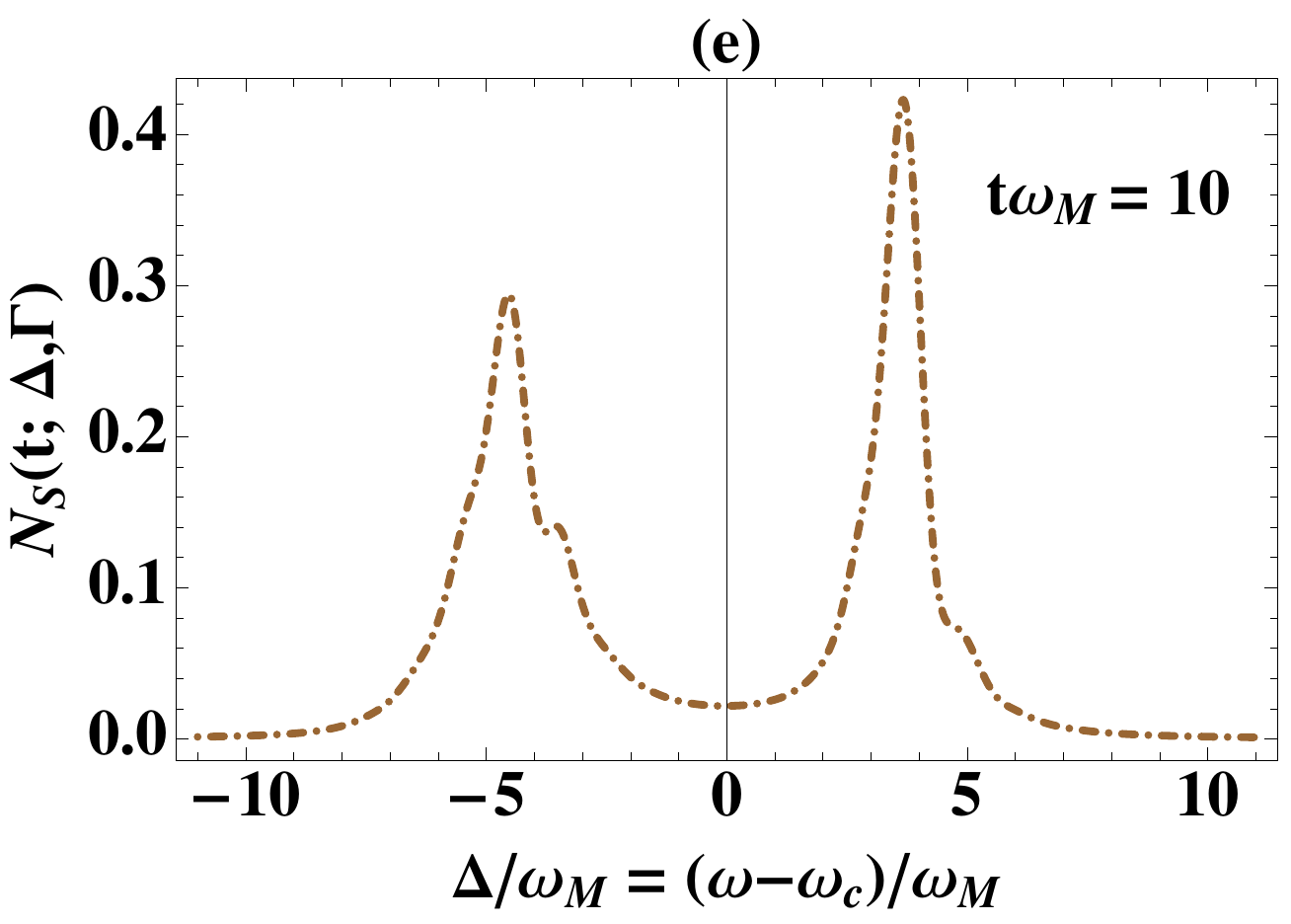} &
    \includegraphics[width=2.25in, height=1.7in]{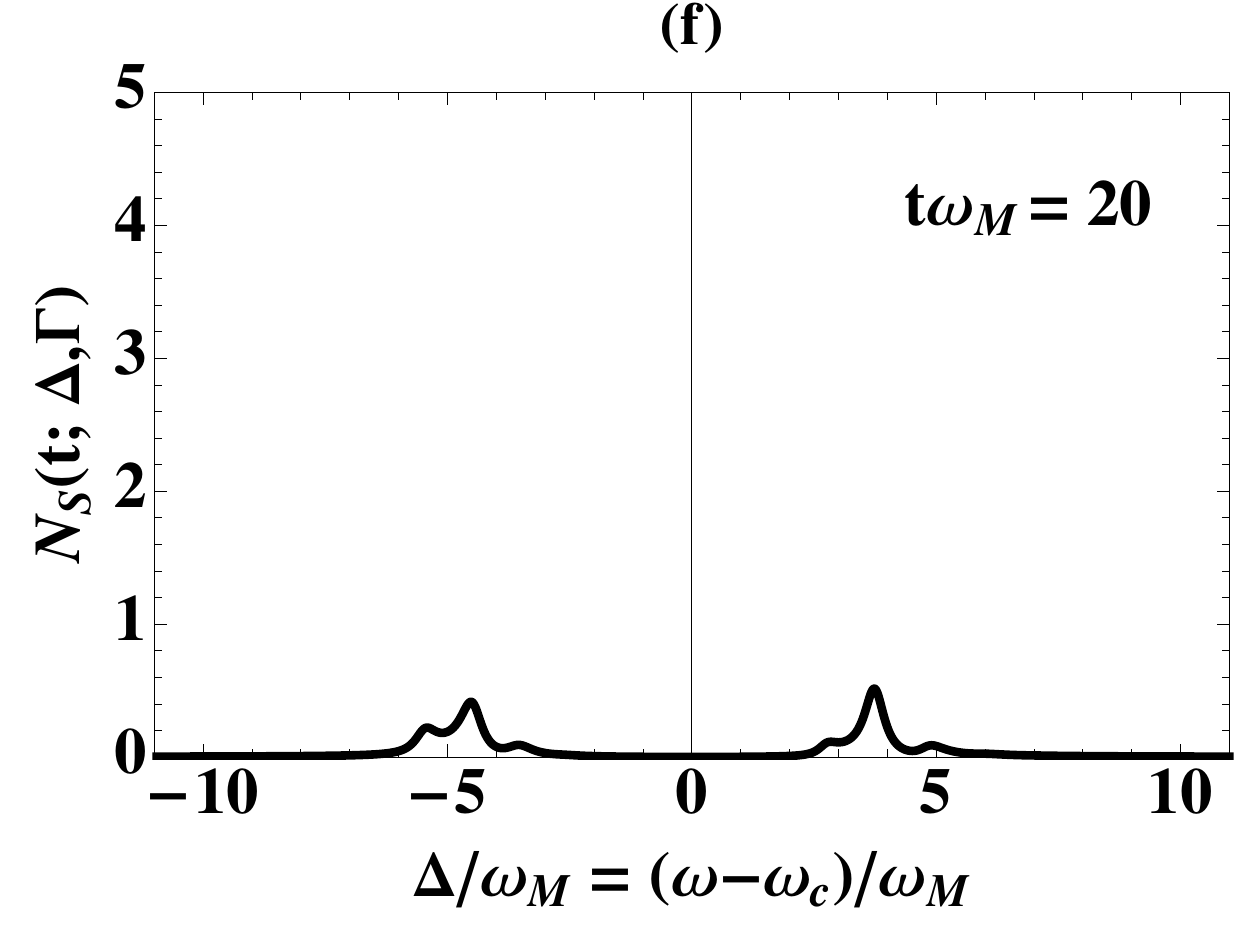} 
  \end{tabular}
  \captionsetup{
  format=plain,
  margin=1em,
  justification=raggedright,
  singlelinecheck=false
}
  \caption{(Color online) Single-photon time-dependent spectrum in a hybrid atom-OMC system, including mechanical damping and spontaneous emission processes. The spectrum is recorded as six different times. $t = 1\omega^{-1}_{M}, 2\omega^{-1}_{M}, 4\omega^{-1}_{M}$, $7\omega^{-1}_{M}, 10\omega^{-1}_{M}$ and $20\omega^{-1}_{M}$ which correspond to thick blue dashed, red dotted, green dotted dashed, thin magenta dashed, brown thin dotted dashed and black solid curves respectively. The parameters used are: $\gamma_{M}/\omega_{M}=0.1,\gamma_{a}/\omega_{M} = 0.4$, $\overline{M} = 0.1$ and rest of the parameters are the same as used in Fig.~2. Note that we have chosen weak damping but a large spontaneous emission loss ($\gamma_{a}\sim\kappa$) in plotting this figure.}
\end{figure*}
In this section we consider a more realistic situation in which two other sources of dissipation: mechanical damping and atomic spontaneous emission are also incorporated in the model. To include mechanical decays we assume the movable mirror is coupled with a finite temperature Markovian mechanical heat bath. The temperature of the bath is characterized by the average thermal phonon number $\overline{M}$ and as a result of this coupling phonons can now leak out and enter into the OMC at a rate $\gamma_{M}$. Assuming the initial state of the mechanical oscillator to be in thermal equilibrium with the mechanical bath, we can express the initial thermal state of the oscillator using the Boltzmann probablity distribution as:
\begin{equation}\label{IBD}
p_{m}(t=0)=\frac{\overline{M}^m}{(1+\overline{M})^{m+1}}.
\end{equation}
while label $m$ is the number of phonons present in the movable mirror. Like previous section, we now apply the quantum trajectory method to calculate the time-dependent spectrum. Non-Hermitian Hamlitonian in this case takes the following form:
\begin{equation}\label{NHHd}
\begin{split}
&\hat{H}_{dNH}=\hat{H}_{sys}-\frac{i\hbar}{2}\hat{J}^{\dagger(O)}\hat{J}^{(O)}-\frac{i\hbar}{2}\hat{J}^{\dagger(A)}\hat{J}^{(A)}\\
&-\frac{i\hbar}{2}(\overline{M}+1)\hat{J}^{\dagger(M)}\hat{J}^{(M)}
-\frac{i\hbar}{2}\overline{M}\hat{J}^{(M)}\hat{J}^{\dagger(M)}
\end{split}
\end{equation} 
with the presence of mechanical bath, the spectrum now become a weighted average over spectra calculated with different initial numbers of phonon states. Alongwith photonic detection events there will be undetected phononic quantum jumps occurring now as well which are represented by jump operator $\hat{J}^{(M)}$ (see Eq.~2(c)).\\

 The jump and no-jump evolution leads to a net decay of phonons from the system which is given by the rate $\gamma_{M}$, as one can also see from the full Master equation analysis of the problem as well. In order to ensure that the optical measurement based state reduction to be a more probable situation than the mechanical decoherence in our setup, we'll also assume a weak mechanical damping limit ($\overline{M}\ll 1$ such that $\kappa >> \gamma_M \overline{M}$) \cite{basiri2012phonon, gangat2011phonon}. Here we would like to emphasize that although quantum trajectory theory can easily be applied to high phonon number case also, but since in present work we are more interested in observable photonic jumps (which will lead us to the optical spectrum) rather than the unobserved mechnaical jumps, hence we are considering a weak damping limit. Notice also that in high damping case, quantum trajectory theory will spend most of its time in calculating the unobserved mechanical quantum jumps, which we want to avoid in present optical spectrum study. This weak damping limit will allow us to neglect the contribution coming from the $\overline{M}$ terms in Eq.(\ref{NHHd}) compared to 1.\\
 
The time-dependent spectrum incorporating phononic and atomic decays is plotted in Fig.~5. Like Fig.~2 we observe that in order to resolve the spectrum into vacumm Rabi spitting it takes some finite time ($\sim t = 4\omega^{-1}_{M}$ as shown in Fig.~5(c)). The novel part of the spectrum is that now the phononic side-bands takes even longer to show up in the spectrum. In the absence of $\gamma_{a}$ and $\gamma_{M}$ we found that the mechanical motion signatures start to appear in the spectrum much earlier ($\sim t = 7\omega^{-1}_{M}$ as shown in Fig.~2(d)), but when phonon losses are taken into account we observe the emergence of phononic side-bands happens later around $ t = 10\omega^{-1}_{M}$. \\
This behavior can be explained based on the fact that it takes some finite time for the single-photon to interact with the movable mirror and to carry the information about this interaction in its spectrum. When there is no mechanical damping present phonons generated due to optomechanical coupling will stay in the OMC and hence photon will exhibit the phonon presence fastly. Compared to that, when phonons can be lost from the OMC quickly, then photon will need more time to show observable effects produced by photon-phonon interaction (phononic side-bands), as one can notice by comparing Fig.~2(d)-(e) and Fig.~5(d)-(e). Similar to Fig.~2, both red and blue side-bands appear at the same time in the spectrum now.\\
Finally at $t=20\omega^{-1}_{M}$ (Fig.~2(f)) we obtain the infinite long time version of the spectrum. We note that although the spectrum has reduced its height considerably, still major vacuum Rabi splitted peaks as well as signs of side-bands are visible. We found also that both mechanical damping and spontaneous emission contribute in the overall peak reduction compared to Fig.~2(f). Through spontaneous emission process this happens due a decrease in the overall photon detection events as considerable number of times the emitted photon is lost into unwanted spontaneous emission channel (a process happening at a rate $\gamma_{a}$) and hence filtered counted rate (time-dependent spectrum) as displayed in Fig.~5(f) shows considerable height reduction compared  to Fig. ~2(f).
Mechanical damping not only causes peak height reduction but also causes a broadening of peaks due to blurring of resonances in the wide thermal background. Note that such type of mechanical blurring effect is also discussed in detail in Ref. \cite{nunnenkamp2011single}, where an empty OMC is driven by a laser field and a cross over from multi-photon to single-photon optomechanics is investigated.

\section{Conclusions}
By employing the quantum jump theory, we have studied the single-photon time-dependent spectrum emitted from a hybrid atom-OMC system in a strong-strong coupling regime. We concluded that the time-dependent spectrum shows how the spectrum builds up in time staring form a broad Lorentzian to a vacuum Rabi splitted frequency doublet and finally showing the emergence of phononic side-bands. Our quantitative analysis provides exact times when optical spectrum suffers these changes. The order in which these resonances appear also describe the time order in which different physical processes become dominant in the system. We found a dressed state picture can be utilized not only to predict the exact locations of different peaks appearing in the spectrum but also to understand unequal peak heights. Finally, by including mechanical losses and spontaneous emission events, we found that heights of all peaks appearing in the spectrum show marked decrease such that tiny side-bands appearing at large $\pm \Delta/\omega_{M}$ values diminish completely.

\bibliography{Article}

\begin{thebibliography}{33}
\expandafter\ifx\csname natexlab\endcsname\relax\def\natexlab#1{#1}\fi
\expandafter\ifx\csname bibnamefont\endcsname\relax
  \def\bibnamefont#1{#1}\fi
\expandafter\ifx\csname bibfnamefont\endcsname\relax
  \def\bibfnamefont#1{#1}\fi
\expandafter\ifx\csname citenamefont\endcsname\relax
  \def\citenamefont#1{#1}\fi
\expandafter\ifx\csname url\endcsname\relax
  \def\url#1{\texttt{#1}}\fi
\expandafter\ifx\csname urlprefix\endcsname\relax\def\urlprefix{URL }\fi
\providecommand{\bibinfo}[2]{#2}
\providecommand{\eprint}[2][]{\url{#2}}

\bibitem[{\citenamefont{Schliesser and
  Kippenberg}(2011)}]{schliesser2011hybrid}
\bibinfo{author}{\bibfnamefont{A.}~\bibnamefont{Schliesser}} \bibnamefont{and}
  \bibinfo{author}{\bibfnamefont{T.~J.} \bibnamefont{Kippenberg}},
  \bibinfo{journal}{Physics} \textbf{\bibinfo{volume}{4}}, \bibinfo{pages}{97}
  (\bibinfo{year}{2011}).

\bibitem[{\citenamefont{Rogers et~al.}(2014)\citenamefont{Rogers, Gullo,
  De~Chiara, Palma, and Paternostro}}]{rogers2014hybrid}
\bibinfo{author}{\bibfnamefont{B.}~\bibnamefont{Rogers}},
  \bibinfo{author}{\bibfnamefont{N.~L.} \bibnamefont{Gullo}},
  \bibinfo{author}{\bibfnamefont{G.}~\bibnamefont{De~Chiara}},
  \bibinfo{author}{\bibfnamefont{G.~M.} \bibnamefont{Palma}}, \bibnamefont{and}
  \bibinfo{author}{\bibfnamefont{M.}~\bibnamefont{Paternostro}},
  \bibinfo{journal}{arXiv preprint arXiv:1402.1195}  (\bibinfo{year}{2014}).

\bibitem[{\citenamefont{Wallquist et~al.}(2010)\citenamefont{Wallquist,
  Hammerer, Zoller, Genes, Ludwig, Marquardt, Treutlein, Ye, and
  Kimble}}]{wallquist2010single}
\bibinfo{author}{\bibfnamefont{M.}~\bibnamefont{Wallquist}},
  \bibinfo{author}{\bibfnamefont{K.}~\bibnamefont{Hammerer}},
  \bibinfo{author}{\bibfnamefont{P.}~\bibnamefont{Zoller}},
  \bibinfo{author}{\bibfnamefont{C.}~\bibnamefont{Genes}},
  \bibinfo{author}{\bibfnamefont{M.}~\bibnamefont{Ludwig}},
  \bibinfo{author}{\bibfnamefont{F.}~\bibnamefont{Marquardt}},
  \bibinfo{author}{\bibfnamefont{P.}~\bibnamefont{Treutlein}},
  \bibinfo{author}{\bibfnamefont{J.}~\bibnamefont{Ye}}, \bibnamefont{and}
  \bibinfo{author}{\bibfnamefont{H.}~\bibnamefont{Kimble}},
  \bibinfo{journal}{Physical Review A} \textbf{\bibinfo{volume}{81}},
  \bibinfo{pages}{023816} (\bibinfo{year}{2010}).

\bibitem[{\citenamefont{Restrepo et~al.}(2014)\citenamefont{Restrepo, Ciuti,
  and Favero}}]{restrepo2014single}
\bibinfo{author}{\bibfnamefont{J.}~\bibnamefont{Restrepo}},
  \bibinfo{author}{\bibfnamefont{C.}~\bibnamefont{Ciuti}}, \bibnamefont{and}
  \bibinfo{author}{\bibfnamefont{I.}~\bibnamefont{Favero}},
  \bibinfo{journal}{Physical review letters} \textbf{\bibinfo{volume}{112}},
  \bibinfo{pages}{013601} (\bibinfo{year}{2014}).

\bibitem[{\citenamefont{Haroche and Kleppner}(1989)}]{haroche1989cavity}
\bibinfo{author}{\bibfnamefont{S.}~\bibnamefont{Haroche}} \bibnamefont{and}
  \bibinfo{author}{\bibfnamefont{D.}~\bibnamefont{Kleppner}},
  \bibinfo{journal}{Phys. Today} \textbf{\bibinfo{volume}{42}},
  \bibinfo{pages}{24} (\bibinfo{year}{1989}).

\bibitem[{\citenamefont{Kavokin et~al.}(2007)\citenamefont{Kavokin, Baumberg,
  Malpuech, and Laussy}}]{kavokin2007microcavities}
\bibinfo{author}{\bibfnamefont{A.}~\bibnamefont{Kavokin}},
  \bibinfo{author}{\bibfnamefont{J.~J.} \bibnamefont{Baumberg}},
  \bibinfo{author}{\bibfnamefont{G.}~\bibnamefont{Malpuech}}, \bibnamefont{and}
  \bibinfo{author}{\bibfnamefont{F.~P.} \bibnamefont{Laussy}},
  \emph{\bibinfo{title}{Microcavities}} (\bibinfo{publisher}{Oxford University
  Press}, \bibinfo{year}{2007}).

\bibitem[{\citenamefont{Kippenberg and Vahala}(2008)}]{kippenberg2008cavity}
\bibinfo{author}{\bibfnamefont{T.~J.} \bibnamefont{Kippenberg}}
  \bibnamefont{and} \bibinfo{author}{\bibfnamefont{K.~J.}
  \bibnamefont{Vahala}}, \bibinfo{journal}{science}
  \textbf{\bibinfo{volume}{321}}, \bibinfo{pages}{1172} (\bibinfo{year}{2008}).

\bibitem[{\citenamefont{Aspelmeyer et~al.}(2014)\citenamefont{Aspelmeyer,
  Kippenberg, and Marquardt}}]{aspelmeyer2014cavity}
\bibinfo{author}{\bibfnamefont{M.}~\bibnamefont{Aspelmeyer}},
  \bibinfo{author}{\bibfnamefont{T.~J.} \bibnamefont{Kippenberg}},
  \bibnamefont{and}
  \bibinfo{author}{\bibfnamefont{F.}~\bibnamefont{Marquardt}},
  \emph{\bibinfo{title}{Cavity Optomechanics: Nano-and Micromechanical
  Resonators Interacting with Light}} (\bibinfo{publisher}{Springer},
  \bibinfo{year}{2014}).

\bibitem[{\citenamefont{Bariani
  et~al.}(2014{\natexlab{a}})\citenamefont{Bariani, Singh, Buchmann,
  Vengalattore, and Meystre}}]{bariani2014hybrid}
\bibinfo{author}{\bibfnamefont{F.}~\bibnamefont{Bariani}},
  \bibinfo{author}{\bibfnamefont{S.}~\bibnamefont{Singh}},
  \bibinfo{author}{\bibfnamefont{L.}~\bibnamefont{Buchmann}},
  \bibinfo{author}{\bibfnamefont{M.}~\bibnamefont{Vengalattore}},
  \bibnamefont{and} \bibinfo{author}{\bibfnamefont{P.}~\bibnamefont{Meystre}},
  \bibinfo{journal}{Physical Review A} \textbf{\bibinfo{volume}{90}},
  \bibinfo{pages}{033838} (\bibinfo{year}{2014}{\natexlab{a}}).

\bibitem[{\citenamefont{De~Chiara et~al.}(2011)\citenamefont{De~Chiara,
  Paternostro, and Palma}}]{de2011entanglement}
\bibinfo{author}{\bibfnamefont{G.}~\bibnamefont{De~Chiara}},
  \bibinfo{author}{\bibfnamefont{M.}~\bibnamefont{Paternostro}},
  \bibnamefont{and} \bibinfo{author}{\bibfnamefont{G.~M.} \bibnamefont{Palma}},
  \bibinfo{journal}{Physical Review A} \textbf{\bibinfo{volume}{83}},
  \bibinfo{pages}{052324} (\bibinfo{year}{2011}).

\bibitem[{\citenamefont{Bariani
  et~al.}(2014{\natexlab{b}})\citenamefont{Bariani, Otterbach, Tan, and
  Meystre}}]{bariani2014single}
\bibinfo{author}{\bibfnamefont{F.}~\bibnamefont{Bariani}},
  \bibinfo{author}{\bibfnamefont{J.}~\bibnamefont{Otterbach}},
  \bibinfo{author}{\bibfnamefont{H.}~\bibnamefont{Tan}}, \bibnamefont{and}
  \bibinfo{author}{\bibfnamefont{P.}~\bibnamefont{Meystre}},
  \bibinfo{journal}{Physical Review A} \textbf{\bibinfo{volume}{89}},
  \bibinfo{pages}{011801} (\bibinfo{year}{2014}{\natexlab{b}}).

\bibitem[{\citenamefont{Yin et~al.}(2013)\citenamefont{Yin, Li, Zhang, and
  Duan}}]{yin2013large}
\bibinfo{author}{\bibfnamefont{Z.-q.} \bibnamefont{Yin}},
  \bibinfo{author}{\bibfnamefont{T.}~\bibnamefont{Li}},
  \bibinfo{author}{\bibfnamefont{X.}~\bibnamefont{Zhang}}, \bibnamefont{and}
  \bibinfo{author}{\bibfnamefont{L.}~\bibnamefont{Duan}},
  \bibinfo{journal}{Physical Review A} \textbf{\bibinfo{volume}{88}},
  \bibinfo{pages}{033614} (\bibinfo{year}{2013}).

\bibitem[{\citenamefont{Kimble}(1998)}]{kimble1998strong}
\bibinfo{author}{\bibfnamefont{H.}~\bibnamefont{Kimble}},
  \bibinfo{journal}{Physica Scripta} \textbf{\bibinfo{volume}{1998}},
  \bibinfo{pages}{127} (\bibinfo{year}{1998}).

\bibitem[{\citenamefont{Brooks et~al.}(2012)\citenamefont{Brooks, Botter,
  Schreppler, Purdy, Brahms, and Stamper-Kurn}}]{brooks2012non}
\bibinfo{author}{\bibfnamefont{D.~W.} \bibnamefont{Brooks}},
  \bibinfo{author}{\bibfnamefont{T.}~\bibnamefont{Botter}},
  \bibinfo{author}{\bibfnamefont{S.}~\bibnamefont{Schreppler}},
  \bibinfo{author}{\bibfnamefont{T.~P.} \bibnamefont{Purdy}},
  \bibinfo{author}{\bibfnamefont{N.}~\bibnamefont{Brahms}}, \bibnamefont{and}
  \bibinfo{author}{\bibfnamefont{D.~M.} \bibnamefont{Stamper-Kurn}},
  \bibinfo{journal}{Nature} \textbf{\bibinfo{volume}{488}},
  \bibinfo{pages}{476} (\bibinfo{year}{2012}).

\bibitem[{\citenamefont{Akram et~al.}(2010)\citenamefont{Akram, Kiesel,
  Aspelmeyer, and Milburn}}]{akram2010single}
\bibinfo{author}{\bibfnamefont{U.}~\bibnamefont{Akram}},
  \bibinfo{author}{\bibfnamefont{N.}~\bibnamefont{Kiesel}},
  \bibinfo{author}{\bibfnamefont{M.}~\bibnamefont{Aspelmeyer}},
  \bibnamefont{and} \bibinfo{author}{\bibfnamefont{G.}~\bibnamefont{Milburn}},
  \bibinfo{journal}{New Journal of Physics} \textbf{\bibinfo{volume}{12}},
  \bibinfo{pages}{083030} (\bibinfo{year}{2010}).

\bibitem[{\citenamefont{Nunnenkamp et~al.}(2011)\citenamefont{Nunnenkamp,
  B{\o}rkje, and Girvin}}]{nunnenkamp2011single}
\bibinfo{author}{\bibfnamefont{A.}~\bibnamefont{Nunnenkamp}},
  \bibinfo{author}{\bibfnamefont{K.}~\bibnamefont{B{\o}rkje}},
  \bibnamefont{and} \bibinfo{author}{\bibfnamefont{S.}~\bibnamefont{Girvin}},
  \bibinfo{journal}{Physical review letters} \textbf{\bibinfo{volume}{107}},
  \bibinfo{pages}{063602} (\bibinfo{year}{2011}).

\bibitem[{\citenamefont{Liao et~al.}(2012)\citenamefont{Liao, Cheung, and
  Law}}]{liao2012spectrum}
\bibinfo{author}{\bibfnamefont{J.-Q.} \bibnamefont{Liao}},
  \bibinfo{author}{\bibfnamefont{H.}~\bibnamefont{Cheung}}, \bibnamefont{and}
  \bibinfo{author}{\bibfnamefont{C.}~\bibnamefont{Law}},
  \bibinfo{journal}{Physical Review A} \textbf{\bibinfo{volume}{85}},
  \bibinfo{pages}{025803} (\bibinfo{year}{2012}).

\bibitem[{\citenamefont{Mirza and van Enk}(2014)}]{mirza2014single}
\bibinfo{author}{\bibfnamefont{I.~M.} \bibnamefont{Mirza}} \bibnamefont{and}
  \bibinfo{author}{\bibfnamefont{S.}~\bibnamefont{van Enk}},
  \bibinfo{journal}{Physical Review A} \textbf{\bibinfo{volume}{90}},
  \bibinfo{pages}{043831} (\bibinfo{year}{2014}).

\bibitem[{\citenamefont{Hennessy et~al.}(2007)\citenamefont{Hennessy, Badolato,
  Winger, Gerace, Atature, Gulde, Falt, Hu, and
  Imamouglu}}]{hennessy2007quantum}
\bibinfo{author}{\bibfnamefont{K.}~\bibnamefont{Hennessy}},
  \bibinfo{author}{\bibfnamefont{A.}~\bibnamefont{Badolato}},
  \bibinfo{author}{\bibfnamefont{M.}~\bibnamefont{Winger}},
  \bibinfo{author}{\bibfnamefont{D.}~\bibnamefont{Gerace}},
  \bibinfo{author}{\bibfnamefont{M.}~\bibnamefont{Atature}},
  \bibinfo{author}{\bibfnamefont{S.}~\bibnamefont{Gulde}},
  \bibinfo{author}{\bibfnamefont{S.}~\bibnamefont{Falt}},
  \bibinfo{author}{\bibfnamefont{E.~L.} \bibnamefont{Hu}}, \bibnamefont{and}
  \bibinfo{author}{\bibfnamefont{A.}~\bibnamefont{Imamouglu}},
  \bibinfo{journal}{Nature} \textbf{\bibinfo{volume}{445}},
  \bibinfo{pages}{896} (\bibinfo{year}{2007}).

\bibitem[{\citenamefont{Cui and Raymer}(2006)}]{cui2006emission}
\bibinfo{author}{\bibfnamefont{G.}~\bibnamefont{Cui}} \bibnamefont{and}
  \bibinfo{author}{\bibfnamefont{M.}~\bibnamefont{Raymer}},
  \bibinfo{journal}{Physical Review A} \textbf{\bibinfo{volume}{73}},
  \bibinfo{pages}{053807} (\bibinfo{year}{2006}).

\bibitem[{\citenamefont{Breyer and Bienert}(2012)}]{breyer2012light}
\bibinfo{author}{\bibfnamefont{D.}~\bibnamefont{Breyer}} \bibnamefont{and}
  \bibinfo{author}{\bibfnamefont{M.}~\bibnamefont{Bienert}},
  \bibinfo{journal}{Physical Review A} \textbf{\bibinfo{volume}{86}},
  \bibinfo{pages}{053819} (\bibinfo{year}{2012}).

\bibitem[{\citenamefont{Jia and Wang}(2013)}]{jia2013single}
\bibinfo{author}{\bibfnamefont{W.}~\bibnamefont{Jia}} \bibnamefont{and}
  \bibinfo{author}{\bibfnamefont{Z.}~\bibnamefont{Wang}},
  \bibinfo{journal}{Physical Review A} \textbf{\bibinfo{volume}{88}},
  \bibinfo{pages}{063821} (\bibinfo{year}{2013}).

\bibitem[{\citenamefont{Jacobs}(2012)}]{jacobs2012probe}
\bibinfo{author}{\bibfnamefont{A.}~\bibnamefont{Jacobs}}, Ph.D. thesis,
  \bibinfo{school}{Miami University} (\bibinfo{year}{2012}).

\bibitem[{\citenamefont{Barclay et~al.}(2009)\citenamefont{Barclay, Santori,
  Fu, Beausoleil, and Painter}}]{barclay2009coherent}
\bibinfo{author}{\bibfnamefont{P.~E.} \bibnamefont{Barclay}},
  \bibinfo{author}{\bibfnamefont{C.}~\bibnamefont{Santori}},
  \bibinfo{author}{\bibfnamefont{K.-M.} \bibnamefont{Fu}},
  \bibinfo{author}{\bibfnamefont{R.~G.} \bibnamefont{Beausoleil}},
  \bibnamefont{and} \bibinfo{author}{\bibfnamefont{O.}~\bibnamefont{Painter}},
  \bibinfo{journal}{Optics express} \textbf{\bibinfo{volume}{17}},
  \bibinfo{pages}{8081} (\bibinfo{year}{2009}).

\bibitem[{\citenamefont{You and Nori}(2011)}]{you2011atomic}
\bibinfo{author}{\bibfnamefont{J.}~\bibnamefont{You}} \bibnamefont{and}
  \bibinfo{author}{\bibfnamefont{F.}~\bibnamefont{Nori}},
  \bibinfo{journal}{Nature} \textbf{\bibinfo{volume}{474}},
  \bibinfo{pages}{589} (\bibinfo{year}{2011}).

\bibitem[{\citenamefont{Dalibard et~al.}(1992)\citenamefont{Dalibard, Castin,
  and M{\o}lmer}}]{dalibard1992wave}
\bibinfo{author}{\bibfnamefont{J.}~\bibnamefont{Dalibard}},
  \bibinfo{author}{\bibfnamefont{Y.}~\bibnamefont{Castin}}, \bibnamefont{and}
  \bibinfo{author}{\bibfnamefont{K.}~\bibnamefont{M{\o}lmer}},
  \bibinfo{journal}{Physical review letters} \textbf{\bibinfo{volume}{68}},
  \bibinfo{pages}{580} (\bibinfo{year}{1992}).

\bibitem[{\citenamefont{Carmichael}(1999)}]{carmichael1999statistical}
\bibinfo{author}{\bibfnamefont{H.~J.} \bibnamefont{Carmichael}},
  \emph{\bibinfo{title}{Statistical methods in quantum optics}},
  vol.~\bibinfo{volume}{2} (\bibinfo{publisher}{Springer},
  \bibinfo{year}{1999}).

\bibitem[{\citenamefont{van Dorsselaer and Nienhuis}(2000)}]{van2000quantum}
\bibinfo{author}{\bibfnamefont{F.~E.} \bibnamefont{van Dorsselaer}}
  \bibnamefont{and} \bibinfo{author}{\bibfnamefont{G.}~\bibnamefont{Nienhuis}},
  \bibinfo{journal}{Journal of Optics B: Quantum and Semiclassical Optics}
  \textbf{\bibinfo{volume}{2}}, \bibinfo{pages}{R25} (\bibinfo{year}{2000}).

\bibitem[{\citenamefont{Eberly and Wodkiewicz}(1977)}]{eberly1977time}
\bibinfo{author}{\bibfnamefont{J.}~\bibnamefont{Eberly}} \bibnamefont{and}
  \bibinfo{author}{\bibfnamefont{K.}~\bibnamefont{Wodkiewicz}},
  \bibinfo{journal}{JOSA} \textbf{\bibinfo{volume}{67}}, \bibinfo{pages}{1252}
  (\bibinfo{year}{1977}).

\bibitem[{\citenamefont{Ren et~al.}(2013)\citenamefont{Ren, Li, Yan, Liu, Xiao,
  and Gong}}]{ren2013single}
\bibinfo{author}{\bibfnamefont{X.-X.} \bibnamefont{Ren}},
  \bibinfo{author}{\bibfnamefont{H.-K.} \bibnamefont{Li}},
  \bibinfo{author}{\bibfnamefont{M.-Y.} \bibnamefont{Yan}},
  \bibinfo{author}{\bibfnamefont{Y.-C.} \bibnamefont{Liu}},
  \bibinfo{author}{\bibfnamefont{Y.-F.} \bibnamefont{Xiao}}, \bibnamefont{and}
  \bibinfo{author}{\bibfnamefont{Q.}~\bibnamefont{Gong}},
  \bibinfo{journal}{Physical Review A} \textbf{\bibinfo{volume}{87}},
  \bibinfo{pages}{033807} (\bibinfo{year}{2013}).

\bibitem[{\citenamefont{Shen and Fan}(2009)}]{shen2009theory}
\bibinfo{author}{\bibfnamefont{J.-T.} \bibnamefont{Shen}} \bibnamefont{and}
  \bibinfo{author}{\bibfnamefont{S.}~\bibnamefont{Fan}},
  \bibinfo{journal}{Physical Review A} \textbf{\bibinfo{volume}{79}},
  \bibinfo{pages}{023837} (\bibinfo{year}{2009}).

\bibitem[{\citenamefont{Basiri-Esfahani
  et~al.}(2012)\citenamefont{Basiri-Esfahani, Akram, and
  Milburn}}]{basiri2012phonon}
\bibinfo{author}{\bibfnamefont{S.}~\bibnamefont{Basiri-Esfahani}},
  \bibinfo{author}{\bibfnamefont{U.}~\bibnamefont{Akram}}, \bibnamefont{and}
  \bibinfo{author}{\bibfnamefont{G.~J.} \bibnamefont{Milburn}},
  \bibinfo{journal}{New Journal of Physics} \textbf{\bibinfo{volume}{14}},
  \bibinfo{pages}{085017} (\bibinfo{year}{2012}).

\bibitem[{\citenamefont{Gangat et~al.}(2011)\citenamefont{Gangat, Stace, and
  Milburn}}]{gangat2011phonon}
\bibinfo{author}{\bibfnamefont{A.~A.} \bibnamefont{Gangat}},
  \bibinfo{author}{\bibfnamefont{T.~M.} \bibnamefont{Stace}}, \bibnamefont{and}
  \bibinfo{author}{\bibfnamefont{G.~J.} \bibnamefont{Milburn}},
  \bibinfo{journal}{New Journal of Physics} \textbf{\bibinfo{volume}{13}},
  \bibinfo{pages}{043024} (\bibinfo{year}{2011}).

\end{thebibliography}
\end{document}